\documentclass[times,twocolumn,final]{elsarticle}
\usepackage{medima}
\usepackage{framed,multirow}
\usepackage{amssymb}
\usepackage{latexsym}
\usepackage{url}
\usepackage{xcolor}
\usepackage[colorlinks=true,linkcolor=blue,citecolor=blue,urlcolor=blue,pagebackref=false]{hyperref}
\definecolor{newcolor}{rgb}{.8,.349,.1}
\journal{XXX}

\usepackage{makecell}
\usepackage{amsmath} 
\usepackage{booktabs}
\newcommand{\tabincell}[2]{\begin{tabular}{@{}#1@{}}#2\end{tabular}}
\usepackage{caption}
\usepackage{multicol}
\usepackage{float}

\begin{document}

\verso{Hao Jiang \textit{et~al.}}

\begin{frontmatter}

\title{Deep Learning for Computational Cytology: A Survey}

\author[1]{Hao Jiang}
\author[2]{Yanning Zhou}
\author[1]{Yi Lin}
\author[3]{Ronald CK Chan}
\author[4]{Jiang Liu}
\author[1]{Hao Chen\corref{cor1}}
\cortext[cor1]{Corresponding author. \\
E-mail address: jhc@cse.ust.hk (Dr. Hao Chen).}
\address[1]{Department of Computer Science and Engineering, The Hong Kong University of Science and Technology, Hong Kong, China}
\address[2]{Department of Computer Science and Engineering, The Chinese University of Hong Kong, Hong Kong, China}
\address[3]{Department of Anatomical and Cellular Pathology, The Chinese University of Hong Kong, Hong Kong, China}
\address[4]{School of Computer Science and Engineering, Southern University of Science and Technology, Shenzhen, China}

\received{XX}
\finalform{XX}
\accepted{XX}
\availableonline{XX}

\begin{abstract}
Computational cytology is a critical, rapid-developing, yet challenging topic in the field of medical image computing which analyzes the digitized cytology image by computer-aided technologies for cancer screening. Recently, an increasing number of deep learning (DL) algorithms have made significant progress in medical image analysis, leading to the boosting publications of cytological studies. To investigate the advanced methods and comprehensive applications, we survey more than 120 publications of DL-based cytology image analysis in this article. We first introduce various deep learning methods, including fully supervised, weakly supervised, unsupervised, and transfer learning. Then, we systematically summarize the public datasets, evaluation metrics, versatile cytology image analysis applications including classification, detection, segmentation, and other related tasks. Finally, we discuss current challenges and potential research directions of computational cytology.

\end{abstract}


\end{frontmatter}


\section{Introduction}\label{Introduction}

\begin{figure*}[t]
	\centering
	\begin{center}
	\includegraphics[width=0.9\textwidth]{{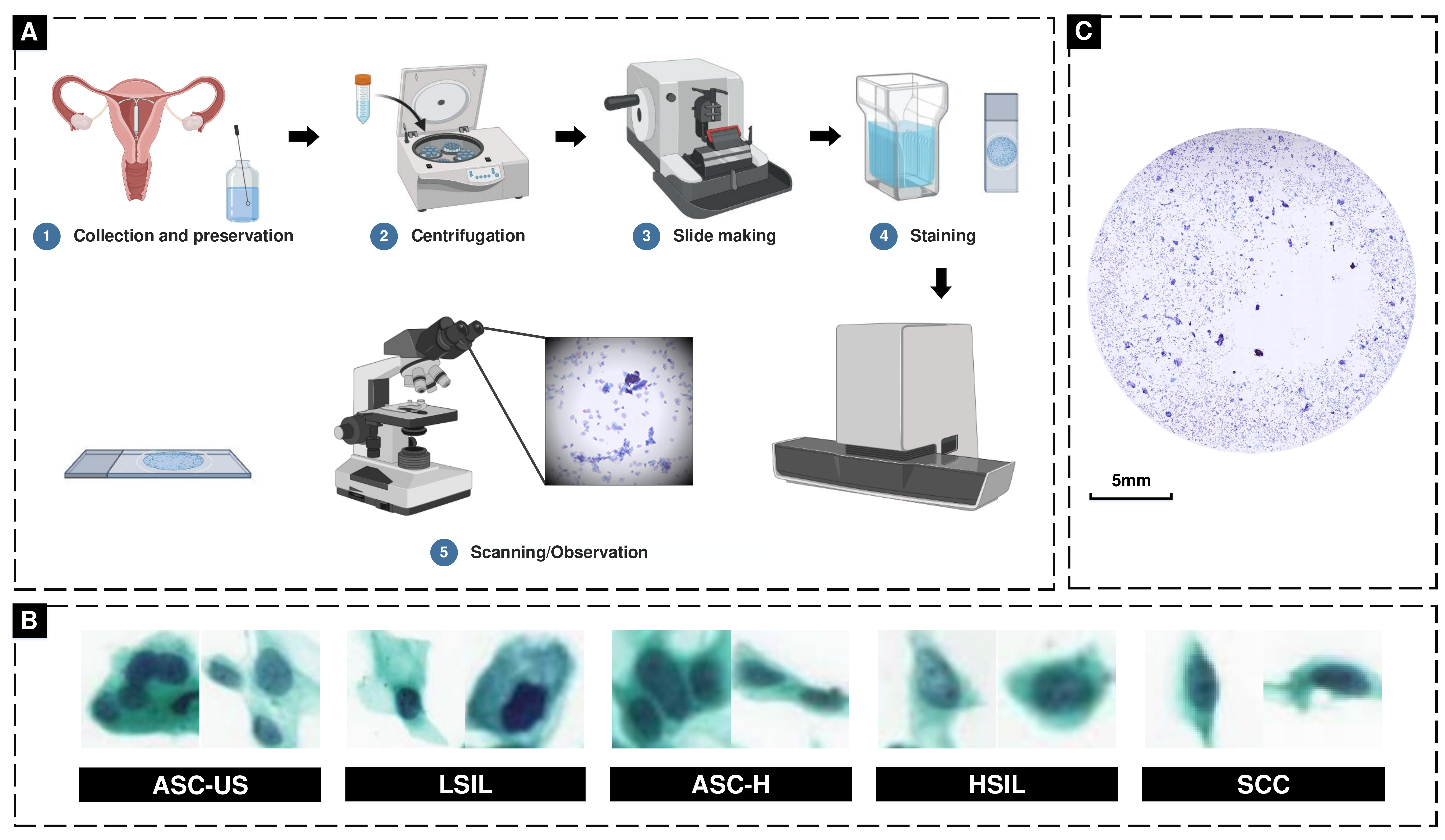}}
	\end{center}
	\caption{Illustration of clinical cytology screening. (A) The procedure of cytology specimen preparation (taking the cervix as an example). (B) Cytology images in different categories. (C) WSI for digital analysis and processing.}
	\label{Fig_1}
	\end{figure*}

Cytology is a branch of pathology to study the cells under microscopes to analyze the cellular morphology, and compositions, usually for cancer screening \citep{davey2006effect,alberts2015essential,o2021diagnostic}.  Compared with histopathology, cytology focuses on the pathological characteristics of cells instead of tissues, which is a collection of thousands of cells in a specific architecture \citep{morrison1993advantages,dey2018basic}. Cells being the structural and functional unit of living organisms \citep{alberts2003molecular}, their morphologies reflect the biological of the organ and even the body \citep{johnston1952cytoplasmic,skaarland1986new,ji2020multimodal}. The clinical cytology testing procedure can be divided into collection and preservation, centrifugation, slide making, and staining (Fig. \ref{Fig_1}(A)). For cytology screening, cytologists observe cytology slides under microscopes and analyze the properties and morphologies of cells (Fig. \ref{Fig_1}(B)). These slides can be also scanned into whole slide images (WSI) (Fig. \ref{Fig_1}(C)) for further digital analysis and processing. In addition, there are three types of cytology specimens, depending on the collection techniques: 1) Exfoliative cytology, including sputum, urine sediment, pleural eﬀusion, and ascites \citep{maharjan2017exfoliative}. 2) Abrasive cytology, including cervical scraping, gastrointestinal tract, and endoscopic brushing \citep{kour2021evaluation}. 3) Aspiration cytology, also named fine-needle aspiration cell inspection (FNAC) \citep{lever1985fine} usually from  breast, thyroid, and lung \citep{koss2006koss,ivanovic2014overview}. Unlike histology, cytology specimens do not require removal of intact sizable tissue, allowing much less invasive sampling procedures. Therefore, the sampling is frequently painless, low-cost, and equipment-undemanding, making cytology useful for cancer screening and early diagnosis \citep{kitchener2006achievements}.

\begin{figure}[t]
		\centering
		\includegraphics[width=0.49\textwidth]{{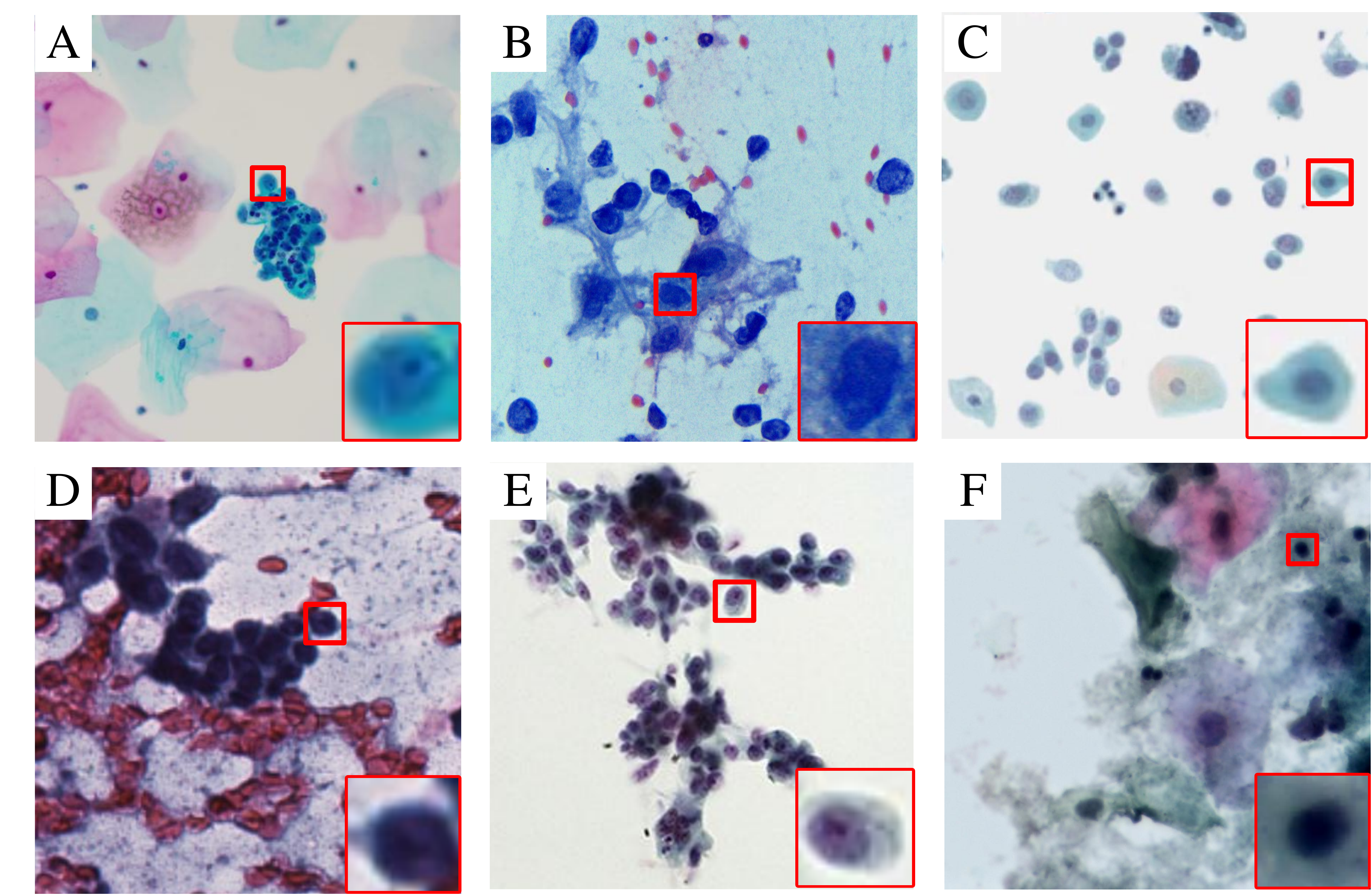}}
		\caption{Typical cytological morphologies of commonly encountered specimens, including (A) Cervix \citep{zhou2019irnet}, (B) Breast \citep{saikia2019comparative}, (C) Urine \citep{awan2021deep}, (D) Thyroid \citep{dov2021weakly}, (E) Lung \citep {teramoto2017automated}, (F) Oral \citep {matias2021segmentation}. Single cellular structures are zoomed by red boxes.}
		\label{Fig_2}
	\end{figure}

Diagnosing cytological specimens is a highly professional task, requiring formal training and assessments overseen by international bodies. Currently, there is a trend to standardize cytological diagnosis reporting, allowing reports and implied risk of cancers to be understood by clinicians without ambiguity \citep{barkan2016paris}. It also provides well-defined features to be looked for among the cells being examined. Fig. \ref{Fig_2} illustrates typical cytological morphologies of commonly encountered specimens. Specifically, cytology screening was ﬁrst applied in cervix cancers almost 100 years ago. Present reporting of cervical cytology is guided by the Bethesda system \citep{nayar2015bethesda}. Pre-cancerous and cancerous cells are first categorized into squamous cells and glandular cells, then they can be identified and graded by combinations of cytological features including enlarged nuclei, multinucleation, perinuclear halo, increased nuclear to cytoplasm ratio, wrinkled nuclear membrane, dyskeratosis (abnormal keratin formation), prominent nucleoli and tumor diathesis. Following the success of Bethesda system, present reporting of breast aspiration cytology is guided by \cite{field2019international}. Aspiration of malignant breast lesions are often hypercellular and the cancer cells possess large sometimes irregular nuclei with prominent nucleoli and the lack of myoepithelial cells. Different grades of atypia were also established to estimate the risk of malignancy \citep{beca2019ancillary}. Similarly, bladder cancer cells from urine often show irregular nuclei with very high N/C ratio ($>$0.7), prominent nucleoli and clumped/coarse chromatin, as outlined in \cite{rosenthal2016paris}. Thyroid cancer aspirates show distinctive features including papillary structures, psammoma bodies and optically clear nuclei with nuclear grooves and nuclear inclusions \citep{cibas20172017}. Together with examples from lung and oral mucosa illustrated in Fig. \ref{Fig_2}, cytological morphology from diﬀerent organs provides diagnostic information for cancer screening and guide patient management at a minimal cost \citep{caddy2005accuracy,kontzoglou2005role}.

In clinical cytology screening, scrutinizing every cell under the microscope (or in gigapixel whole slide images) in search of malignancy can be very time-consuming and tedious for cytologists \citep{mehrotra2011efficacy, de2014large}. Considering ever increase in caseload, researchers have attempted to develop automatic methods for accurate and eﬃcient cancer screening. The ﬁrst successful trial could date back to the 1950s when an automatic screening system was developed for cervix \citep{tolles1956automatic, tolles1955section}. Afterwards, a series of cervical screening systems were launched with varying market success \citep{koss1994evaluation, wilbur2009becton, brown1999cost}. These automated cytology screening systems have been shown to improve the eﬃciency without compromising accuracy of cytology screening procedures.

In recent decades, the automation technology and artiﬁcial intelligence (AI) have achieved remarkable progress in the ﬁeld of medicine. Machine learning (ML), which is a subfield of AI, focuses on learning algorithms to represent the underlying patterns of data by imitating human beings. With rapid progress in ML, medical image interpretation and computer assisted-diagnosis in pathology (e.g., histopathology, cytology \citep{chen2016mitosis,zhao2019pfa,zhu2021hybrid}), and radiology (e.g., computed tomography (CT), magnetic resonance imaging (MRI), X-ray, ultrasound \citep{lassau2021integrating,jonsson2019brain,ccalli2021deep}). For cytology, previous studies demonstrated the feasibility of various ML approaches in cytology image analysis, including support vector machine (SVM), fuzzy c-means (FCM) \citep{chankong2014automatic}, k-means \cite{isa2005automated}, and fuzzy clustering \citep{plissiti2009automated}. However, there remains challenges in these machine learning algorithms, such as developing accurate and efficient cytology image analysis approaches, establishing human-machine collaborative cytological screening systems.

Deep learning, as a branch of the ML family, was developed with multilayers neural networks for leveraging feature representations of input data. DL aims to reduce the heavy reliance on task-related features designed from expert knowledge in traditional ML approaches. It can also increase the model's capability of feature representation by end-to-end learning. In computational cytology, DL could provide cytologists with feasible solutions for accurate and efficient cytological screening. These approaches have been widely investigated in versatile types of cancers, such as cervix \citep{rahaman2020survey}, breast \citep{garud2017high}, bladder \citep{dov2021weakly}, and lung \citep{teramoto2017automated}. Among these DL-based methods, supervised learning involves mapping input images to predeﬁned labels and it has been the most commonly developed DL scheme. Most existing studies on cytological applications focus on improving DL models performance by introducing specific constrains or architecture designs of DL models, such as introducing morphological constraints \citep{zhou2019cia,chai2021deep,chen2017dcan}.

\begin{figure}
		\centering
		\includegraphics[width=0.49\textwidth]{{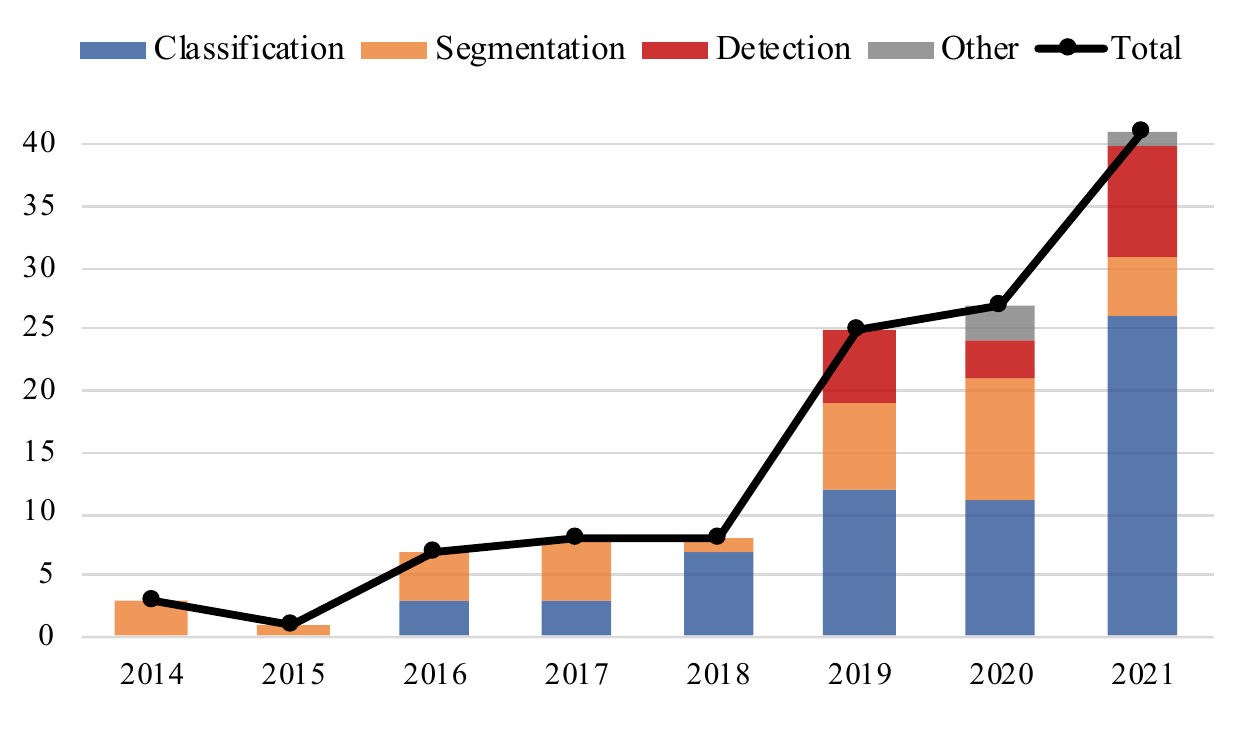}}
		\caption{Number of publications in deep learning-based computational cytology of classification, detection, segmentation and other tasks.}
		\label{Fig_3}
	\end{figure}

The advancement of DL has greatly accelerated the development of computational cytology. There is a 10-fold increase in DL-based computational research from 2014 to 2021 with a booming trend in recent two years. The number of related publications is illustrated in Fig.~\ref{Fig_3}, after searching literature databases (Google Scholar, PubMed, and arXiv). These publications focus on developing various DL approaches for cytological screening, such as classifying between normal and abnormal cells  \citep{zhang2017deeppap}, locating and identifying cells in cytological smears \citep{pirovano2021computer}, and segmentation of different cellular compartments \citep{zhou2020deep}. 

There exists several surveys in the field of cytology image analysis \citep{landau2019artificial,mitra2021cytology, rahaman2020survey}. However, these reviews were far from exhaustive in terms of the advanced algorithms, publicly available datasets, and promising trends in this field. Besides, most of the DL-based cytology surveys focused on the cervix, ignoring the progress in other types of cancer, such as lung and bladder \citep{rahaman2020survey}. Afterwards, \cite{mitra2021cytology} focused on various cytology applications instead of analyzing them in the DL methodology perspective. In this paper, we have surveyed over 120 publications since 2014, and systematically reviewed the progress of DL approaches and techniques in computational cytology, also covering cytology specimens from various parts of the body.

There are six sections in this paper. \textbf{Section \ref{Introduction}} briefly introduces the background and objective of this review. \textbf{Section \ref{Deep learning methodology}} gives an overview of different learning approaches in the context of computational cytology. \textbf{Section \ref{Datasets and metrics}} summarizes public cytology datasets and common evaluation metrics. \textbf{Section \ref{Deep learning in cytology application}} presents the progress and achievements on the DL-based cytology image analysis. \textbf{Section \ref{Promises and Challenges}} discusses existing challenges and potential research directions in computational cytology. \textbf{Section \ref{Conclusion}} concludes this survey paper.

\section{Deep learning methodology} \label{Deep learning methodology}
In this section, we present the definition, formulations, and general procedures of DL methods, which can be developed for various cytological applications. According to the availability of annotations, DL can be categorized as supervised learning (section \ref{Supervised learning}), weakly supervised learning (section \ref{Weakly supervised learning}), unsupervised learning (section \ref{Unsupervised learning}), together with transfer learning (section \ref{Transfer Learning}).

	\begin{figure*}[t]
	\begin{center}
	\includegraphics[width=0.9\textwidth]{{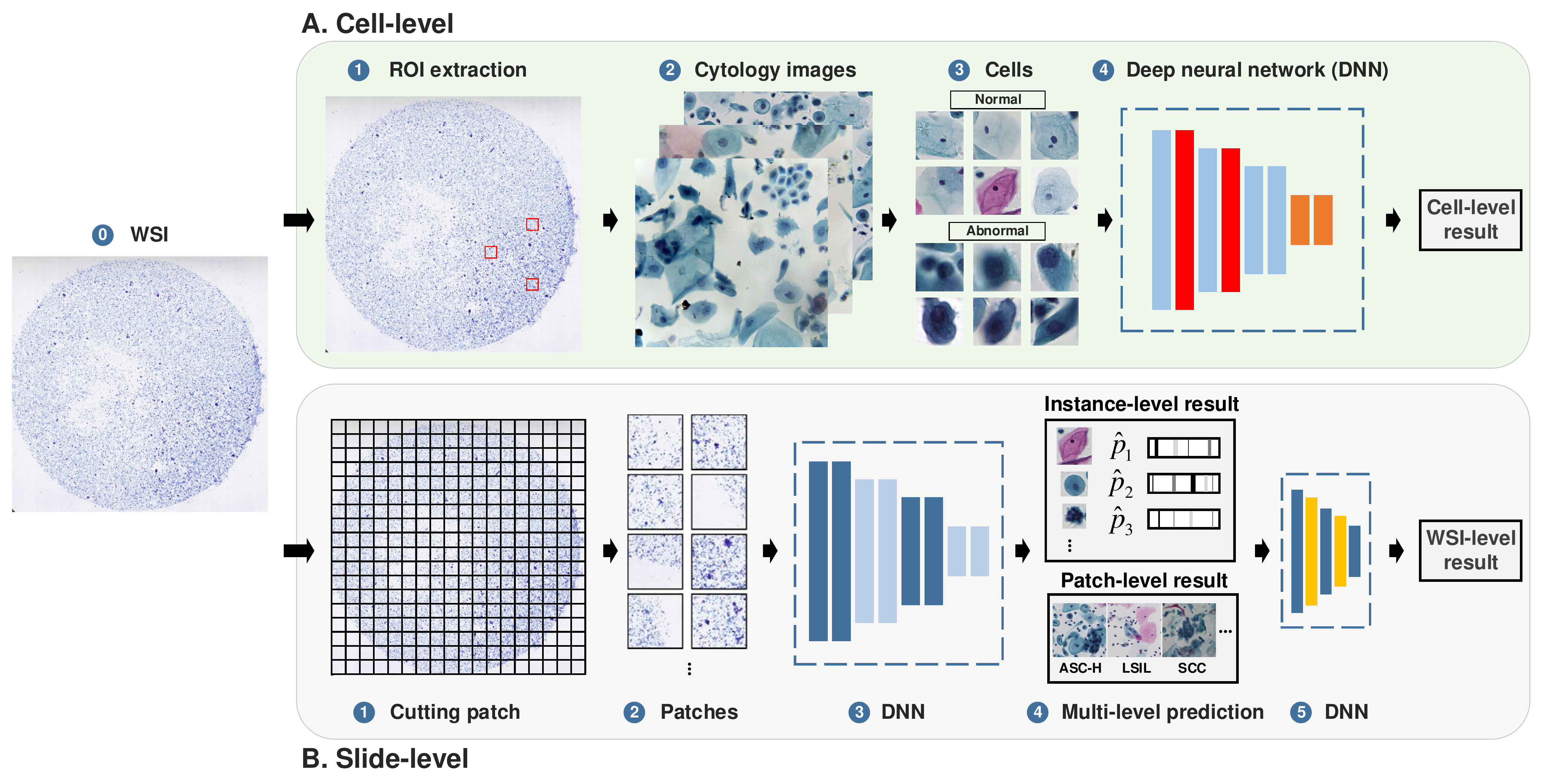}}
	\end{center}
	\caption{Standard workflow of DL-based supervised learning for cytological classification. (A) Cell-level: 
Firstly, ROIs are extracted from WSIs and cut into cell patches, then they are input into the DNN for extracting features and predicting category of each cell patch. (B) Slide-level: Patches cut from WSIs are input into a DNN to obtain multi-level predictions (e.g., instance-level and patch-level), then these predictions are used to predict final WSI-level results.}
	\label{Fig_4}
	\end{figure*}
	
\subsection{Supervised learning} \label{Supervised learning}
Supervised learning aims to learn functional mappings between input data and corresponding labels. For medical image analysis, the inputs are medical images, while labels are varied according to different tasks, e.g., image-level categories for classification, object-level localizations (e.g., boxes, points) for detection, and pixel-wise masks for segmentation. Formally, the input images $X=\{x_{i}\}_{i=1}^{N} $ together with corresponding labels $Y=\{y_{i}\}_{i=1}^{N} $ are used to train a predictive model by minimizing the objective function. The typical deep models in supervised learning include multilayer perceptron (MLP) \citep{rosenblatt1961principles}, convolutional neural network (CNN) \citep{lecun1998gradient}, recurrent neural network (RNN) \citep{zaremba2014recurrent}, and transformer \citep{tay2020efficient}.

CNN is regarded as the most successful DL architecture in image analysis \citep{litjens2017survey}. It mainly consists of three types of hidden layers: convolutional layers for feature extraction, pooling layers for reducing the feature resolution, and fully connected layers for compiling the features extracted by previous layers and outputting prediction results. Then, backpropagation algorithm is introduced to update parameters of different layers during training \citep{lecun1998gradient}. Due to the strategies of local receptive fields, shared weights, and downsampling in pooling layers, CNN has achieved great success in many image analysis tasks, such as autonomous driving \citep{rosenzweig2015review}, face recognition \citep{wang2021deep}, and biomedicine \citep{litjens2017survey}.

Commonly-used CNN architectures in computer vision fields have been employed and developed for various cytological applications, e.g., AlexNet \citep{mohammed2021single}, VGGNet \citep{albuquerque2021ordinal}, ResNet \citep{miselis2019deep}. Currently, most cytology researches focus on developing new algorithms based on these basic architectures in various DL tasks: classification, detection, and segmentation \citep{lin2019fine,sornapudi2019comparing,hussain2020comprehensive}.

\noindent
\textbf{Classification.} This classification model aims at predicting the category of the input image. It can be essentially formulated as $\hat{y}=f(x, \theta)$, where $x$ and $\hat{y}$ is input image and its predicted category, and $\theta$ represents learnable parameters of classification architecture. For training these architectures, cross-entropy loss $L_{CE}$ measures the discrepancy between predicted probability $\hat{y_{ic}}$ and true label $y_{ic}$ by probability distribution:
\begin{equation}
L_{C E}=-\frac{1}{N} \sum_{i=1}^{N} \sum_{c=1}^{M}y_{ic} \log \left(\hat{y}_{ic}\right)
\end{equation}
where, $N$ is the amount of image samples, and $M$ is the number of categories. Then, the prediction loss is used to optimize parameters $\theta$ of network by backpropagation \citep{rumelhart1986learning}. 
	
Usually, the label space of the cytology image refers to its benign/malignant or sub-category. In the standard cell-level classification workflow, regions of interest (ROIs) are extracted from collected slides by cytologists or technicians. Then, ROIs are cut into cell patches as input of deep models. After that, a DL-based feature extractor is responsible for representing high-level features and outputting prediction categories (Fig. \ref{Fig_4}). For slide-level screening, existing studies mainly divide this task into two stages: First, the deep neural network (e.g., CNN, RNN, and Transformer) is responsible for predicting multi-level results, such as detection of malignant or benign cells and the category of patches. Then, another network aggregates these results and predicts the final WSI-level results \citep{wei2021efficient, lin2021dual}.

\noindent
\textbf{Detection.} Unlike classification, the detection task is to locate objects from whole images and predict categories of these objects. Thus, it can be regarded as the combination of two tasks: regressing the object's location and classifying the types of objects. CNN-based object detection algorithms are mainly divided into two categories: two-stage method and one-stage method. In two-stage, the workflow includes feature extraction, region proposal, and prediction. The first stage is to regress coarse prediction (box location and predicted probability) by region proposal. The second stage aims to output fine predictions (box location and object category). Typical models for two-stage algorithms include R-CNN \citep{girshick2014rich}, Fast R-CNN \citep{girshick2015fast}, Faster R-CNN \citep{ren2015faster}. One-stage detection methods aim to abandon the strategy of region proposal. Instead, they directly predict the category and location of the objects in an end-to-end architecture, including SSD \citep{liu2016ssd}, YOLO \citep{redmon2016you}, FCOS \citep{tian2019fcos} and RetinaNet \cite{lin2017focal}.

There are two types of loss functions in object detection task, classification loss and location loss. For classification loss, by improving $L_{CE}$, focal loss $L_{focal}$ is proposed in RetinaNet to balance classification samples \citep{lin2017focal}:
\begin{equation}
L_{focal}=\left\{\begin{array}{ccc}
-\alpha(1-p)^{\gamma} \log (p), & \text {if} & y=1 \\
-(1-\alpha)p^{\gamma} \log (1-p), & \text{if} & y=0
\end{array}\right.
\end{equation}
where $p$ is the model’s estimated probability with label $y$, and $\gamma$ and $\alpha$ are tunable parameters. For location loss, mean absolute error loss $L_{MAE}$ calculates the average distance between the predicted and the true locations:
\begin{equation}
L_{MAE}=\frac{\sum_{i=1}^{N}\left|f\left(x_{i}\right)-y_{i}\right|}{N}
\end{equation}
Then, intersection over union (IoU) loss was introduced to calculate the loss of predicted boxes instead of coordinates in $L_{MAE}$ \citep{yu2016unitbox}:
\begin{equation}
L_{IoU} =-\ln \frac{\text {Intersection}\left(box_{gt}, box_{p}\right)}{\text { Union }\left(box_{gt}, box_{p}\right)}
\end{equation}
where $box_{gt}$ and $box_{p}$ represent ground truth box and predicted box, respectively. After that, some advanced loss functions are designed recently, including GIoU loss for non-intersection area \citep{rezatofighi2019generalized}, CIoU loss for closing center points \citep{zheng2020distance}, etc. 

In addition, there are still someFor example, fully common issues for both one-stage and two-stage algorithms. For example, multiple overlapping predicted boxes in the prediction results. These repetitive and redundant proposals can be removed by the strategy of non-maximum suppression \citep{neubeck2006efficient}.

\begin{figure}[t]
		\centering
		\includegraphics[width=0.49\textwidth]{{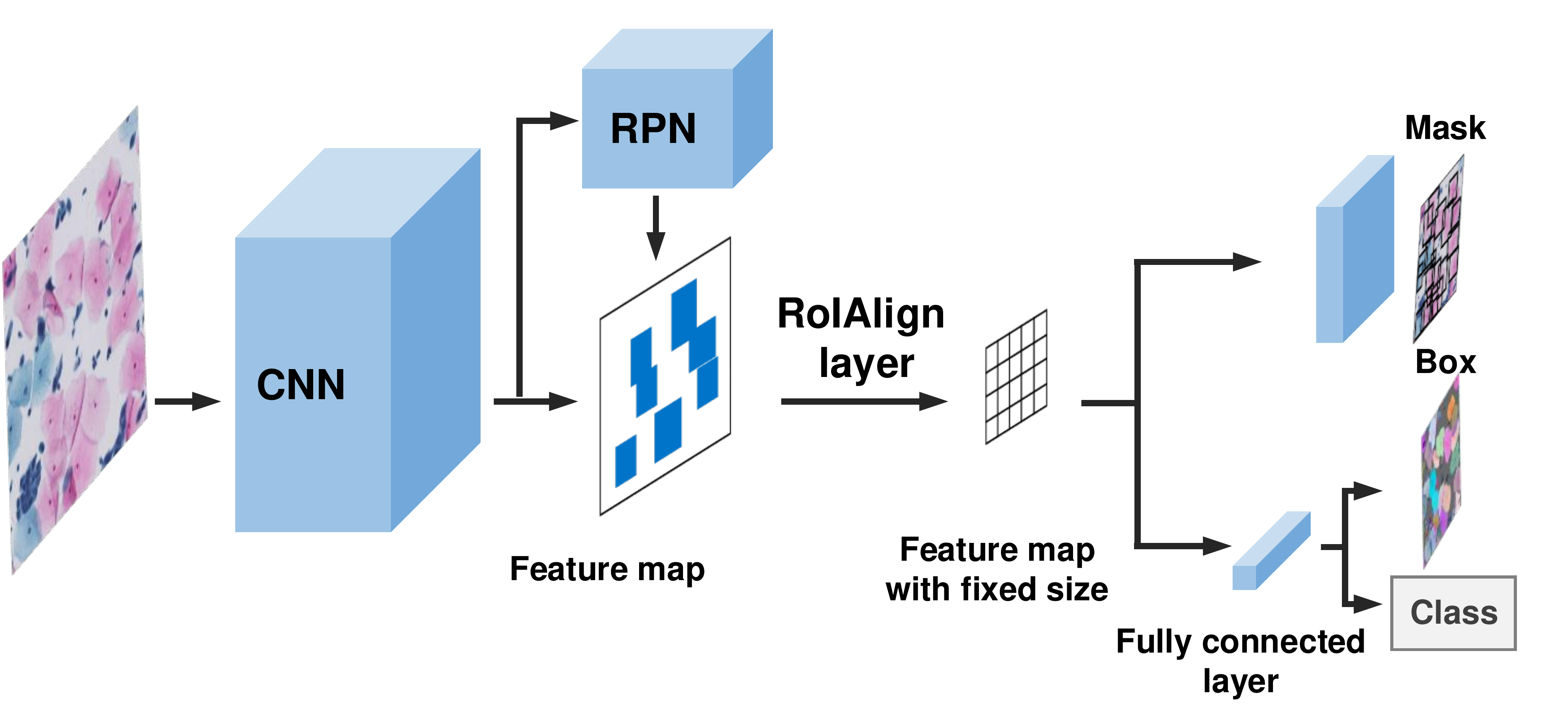}}
		\caption{Instance segmentation in cytology image analysis by Mask R-CNN.}
		\label{Fig_5}
	\end{figure}

\noindent
\textbf{Segmentation.} This is a fundamental and essential task in medical image analysis. Segmentation is to make the pixel-wise prediction which represents the morphology of biomedical structures, such as cell \citep{zhou2020deep}, gland \citep{chen2016dcan}, and organ \citep{rahaman2020survey}. According to whether to distinguish each instance object, DL-based segmentation models can be divided into two branches: semantic segmentation and instance segmentation. Semantic segmentation aims to predict the category of each pixel to obtain masks of objects, which can be regarded as a pixel-wise classification task. Fully convolutional network (FCN) is one of the successful segmentation architectures, which replaced the fully connected layers in traditional CNN with convolutional layers for outputting segmentation map \citep{long2015fully}. Then, \cite{falk2019u} proposed U-Net for biomedical image segmentation by multi-scale feature fusion with a downsampling-upsampling architecture. In instance segmentation, models not only segment pixels into categories but also assigned them corresponding instance ID. Its popular structures mainly follow the detect-then-segment pipeline (Fig. \ref{Fig_5}). For instance, Mask R-CNN was proposed by introducing a mask prediction head based on Faster R-CNN \citep{he2017mask}.

When training mentioned semantic and instance segmentation models, classification loss $L_{CE}$ is introduced as the pixel-wise classification supervision for segmentation. Another widely-used loss function, dice coefficient loss $L_{dice}$ is designed to measure the similarity between predicted masks $Y_{m}$ and ground truth $Y_{gt}$ by calculating the dice coefficient, and is defined as \citep{milletari2016v}:
\begin{equation}
L_{dice}=1-\frac{2|Y_{m} \cap Y_{gt} |}{|Y_{m} |+|Y_{gt} |}
\end{equation}

\subsection{Weakly supervised learning}\label{Weakly supervised learning}
Weakly supervised learning is proposed for the scenarios where labels are not fully available, including incomplete supervision, inaccurate supervision, and inexact supervision \citep{zhou2018brief}. 
	\begin{figure*}[t]
	\begin{center}
	\includegraphics[width=0.95\textwidth]{{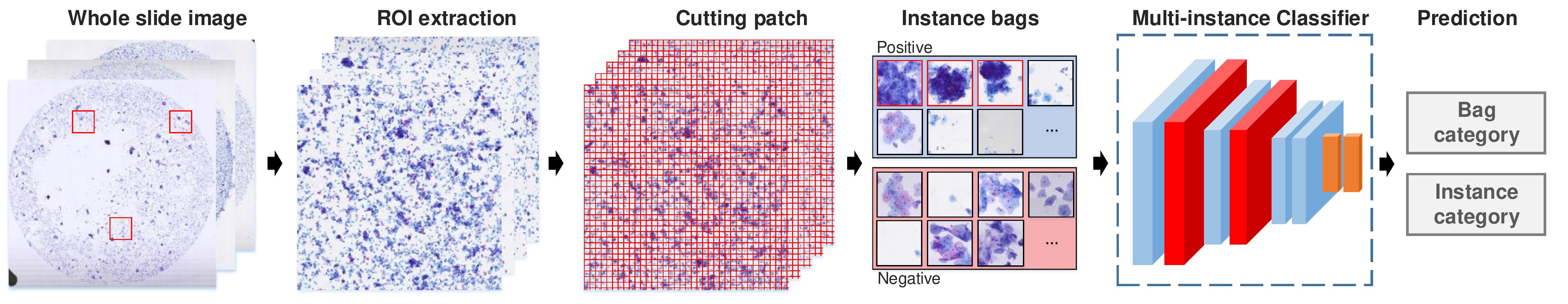}}
	\end{center}
	\caption{Overview of multiple instance learning paradigm. Original WSIs are extracted to ROIs and cut into patches, then they are formed instance bags with bag-level label (positive or negative). For learning algorithms, these bags are used to learn a classification function that can predict the labels of bags and instances in the testing data.}
	\label{Fig_6}
	\end{figure*}
Semi-supervised learning is one successful learning paradigm of incomplete supervision. It leverages both labeled and unlabeled data by extracting hidden information in unlabeled sets to enhance the feature representation of the labeled set. For inaccurate supervision, methods for noisy label problem can identify the potentially mislabeled samples and make corrections, thus improving the reliability of supervision.

Multi-instance learning (MIL) is an effective inexact supervision method, which aims at utilizing coarse-level annotations (e.g., image-level) for learning fine-level (e.g., pixel-level, patch-level) tasks \citep{maron1998framework,xu2014deep}. The standard workflow of MIL is illustrated in Fig. \ref{Fig_6}. Firstly, a series of patches are extracted from whole images with patch-level annotations. Then, these patches are cut into instances and formed bags. Finally, a multi-instance classifier is established by learning for multi-instance bags, which is used for the prediction of unknown bags or instances. Specifically, given a training dataset $\{ (X_{i}, Y_{i})_{i=1}^{N}\}$, where $ X_{i}=\left\{x_{i1}, x_{i2}, \ldots, x_{i, m}\right\}$ are instance bags, $x_{i, m}$ is the $m\text{-}th$ instance of the $i\text{-}th$ bag. $Y_{i} \in\{-1,+1\}$ is its corresponding label of the $i\text{-}th$ bag, +1 represents positive bag with at least one positive instance in this bag, and -1 represents negative bag with no positive instance. Then, bags $X_{i}$ together with labels are used to learn a classification function that can predict the labels of bags and instances.

Weakly supervised learning is particularly appealing in computational cytology scenarios (e.g., whole slide thyroid malignancy prediction \citep{dov2021weakly}), where full labels are expensive to obtain \citep{srinidhi2020deep}. Because fully labeling of all lesions and cells in cytological screening WSIs is hardly possible for cytologists. Hence, weakly supervised learning is introduced to effectively represent and enhance features in the scenarios of limited annotations.

\subsection{Unsupervised learning}\label{Unsupervised learning}
Unsupervised learning is effective for learning useful and underlying representations from unlabeled data, which can be utilized for downstream tasks. For example, unsupervised image augmentation can increase the amount and variety of original dataset for increasing the performance of classification models. Afterwards, unsupervised stain transformation can be adopted to normalize datasets in preprocessing pathology images. Auto-encoder (AE) is a typical structure in unsupervised learning, which is formulated as: $\mathrm{P}(x_{i}) \rightarrow z \rightarrow \mathrm{P}\left(x_{i}^{\prime}\right)$, where AE is trained to encode the input image $ x_{i}$ to obtain latent representation $z$. Then, decoders generate reconstructed image $ x_{i}^{\prime}$ with the supervision of the raw input $x_{i}$. 

\begin{figure}[t]
		\centering
		\includegraphics[width=0.49\textwidth]{{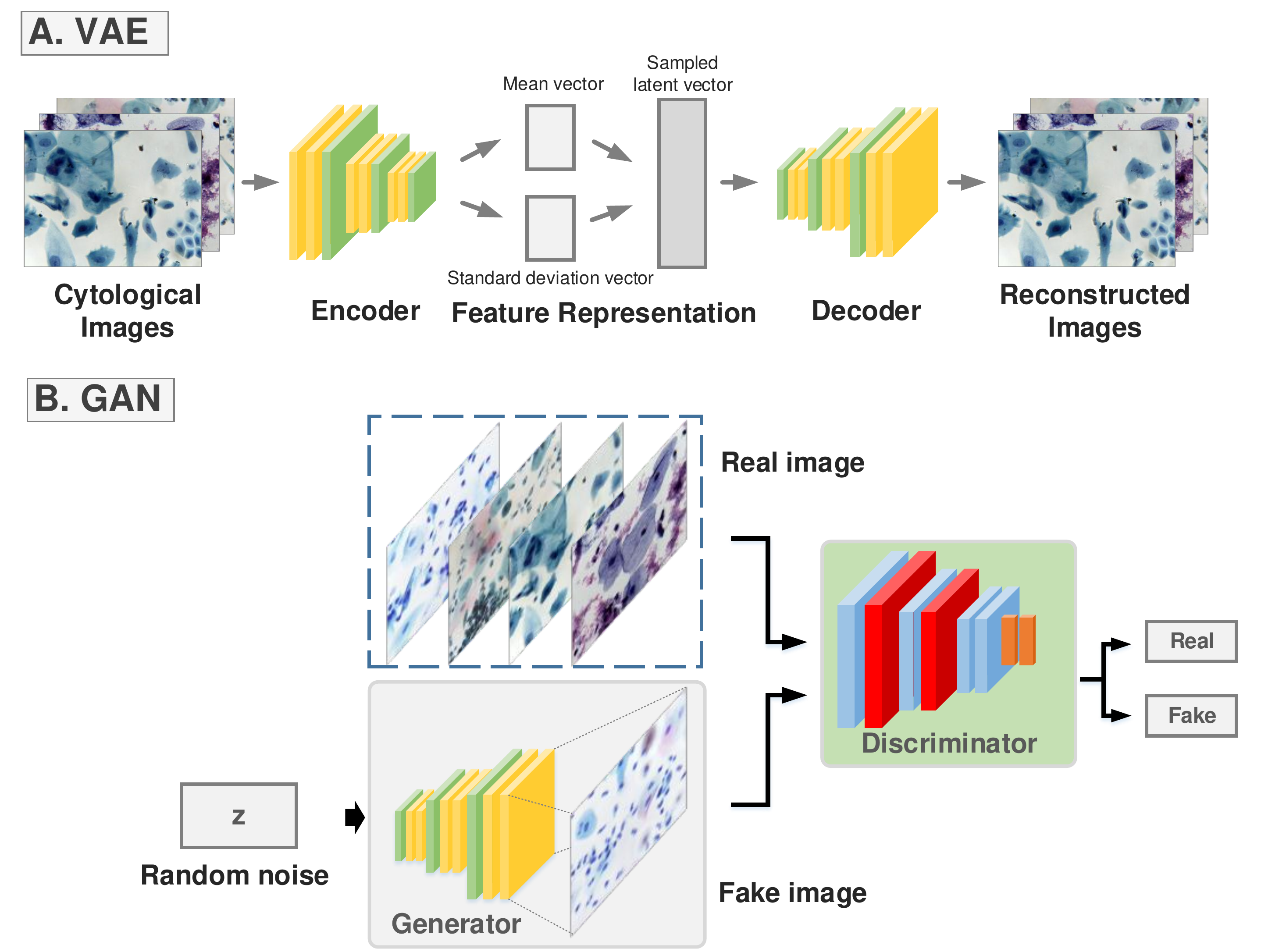}}
		\caption{AE-based unsupervised learning models. (A) Variational auto-encoder (VAE) improved AE by constraining the latent representation to be the normal distribution. (B) Generative adversarial network (GAN), which consists of a generator and a discriminator. The generator is responsible for generating fake images from random noise while discriminator forcing the generator generating realistic images by discriminating generated fake image and real image.}
		\label{Fig_7}
	\end{figure}

Two typical unsupervised models have gained popularities: variational auto-encoder (VAE) \citep{larsen2016autoencoding,kingma2013auto} and generative adversarial network (GAN) \citep{goodfellow2014generative}. As shown in Fig. \ref{Fig_7}(A), VAE improved AE by constraining latent variables to be normally distributed, then sampling a latent vector into a decoder for outputting the image. GAN is another promising architecture that can mitigate the difficulty of collecting large-scale labeled medical datasets by synthesizing high-quality fake images. For its structure, GAN is formed by a generator-discriminator architecture, as shown in Fig. \ref{Fig_7}(B). The generator aims to generate realistic images, while the discriminator competes with generator to differentiate real images and generated images. Therefore, this generator-discriminator architecture is optimized via the adversarial training \citep{goodfellow2014generative}:
\begin{equation}
\begin{split}
\min _{G} \max _{D} V(D, G)=\mathbb{E}_{x \sim p_{\text {data }}(x)}[\log D(x)] \\
 +\mathbb{E}_{z \sim p_{\text {z }}(z)}[\log (1-D(G(z)))]
 \end{split}
\end{equation}
where $G (\cdot)$ denotes the generator, and $D (\cdot)$ denotes the discriminator. $\mathbb{E(\cdot)}$ is the expectation value of distribution, $p_{\text {data }}(x)$ and $ p_{\text {z }}(z)$ are the distribution of the real sample and noise, respectively. During training, generator $G (\cdot)$ aims to learn the distribution of real samples $p_{\text {data}}$, and discriminator $D (\cdot)$ is responsible for discriminating generated and real images, thus forcing the generator to generate realistic images. After the emergence of this basic GAN structure and adversarial loss, many advanced GAN models are proposed to satisfy higher requirements of generated images (e.g., quality, fidelity, and diversity) \citep{yi2019generative}. For example, one promising architecture, CycleGAN is designed for style transformation between unpaired images ($s$, $t$), which is a symmetrical structure consisting of two generators $\{G_{S \rightarrow T}, G_{T \rightarrow S}\}$ for mutual generation between two domains ($S$ and $T$), and two discriminators $\{D_{S}, D_{T}\}$ for discriminating generated images of respective domains. In addition, cycle consistency loss $L_{cyc}(G_{S \rightarrow T}, G_{T \rightarrow S}) $ is designed for one-to-one mapping in CycleGAN architecture, defined as \citep{zhu2017unpaired}:
\begin{equation}
\begin{split}
L_{cyc}(G_{S \rightarrow T}, G_{T \rightarrow S}) =\mathbb{E}_{s \sim p_{\text {data }}(x)}\left[\|G_{T \rightarrow S}(G_{S \rightarrow T}(s))-s\|_{1}\right] \\
 +\mathbb{E}_{t \sim p_{\text {data }}(y)}\left[\|G_{S \rightarrow T}(G_{T \rightarrow S}(t))-t\|_{1}\right]
\end{split}
\end{equation}
where $p_{\text{data}}(s)$ and $p_{\text{data}}(t)$ are the distribution of images in domain $S$ and domain $T$.

In cytology, unsupervised learning algorithms have been designed for various DL tasks, such as stain conversion by CycleGAN \citep{teramoto2021mutual}, data augmentation for improving the accuracy of classification by cGAN \citep{dey2019syncgan}, and generating high-resolution images by GAN-based super-resolution model \citep{ma2020pathsrgan}.

\subsection{Transfer Learning}\label{Transfer Learning}
Transfer learning is a subfield of deep learning that focuses on transferring knowledge from source to target domain for enhancing target tasks. Two transfer learning approaches are commonly used in medical image analysis, including fine-tuning and domain adaptation (DA).

Fine-tuning is regarded as a common model initialization trick for training DL models. It can reduce the overfitting issue and improve the generalization capability of deep models by transferring knowledge from large public datasets (e.g., ImageNet \citep{deng2009imagenet}) to domain-specific tasks (e.g., cervical cell classification) \citep{sornapudi2019comparing,yosinski2014transferable}. Formally, the goal of fine-tuning is training a task $\mathcal{T}^{t}$ by a small dataset $T$. Specifically, it exploits a large-scale and task-similar $\mathcal{T}^{t}$ dataset $S=\{s_{i}\}_{i=1}^{M} (M>>N) $ to pre-train a network $f(\sim;\theta)$ first, then a small target dataset $T=\{t_{i}\}_{i=1}^{N} $ is used to train the several last layers of the pre-trained model for obtaining the target model $N_{t}$. Under the premise that different datasets for training similar tasks have similar low-level feature representations, fine-tuning is regarded as a common and effective training strategy in various DL tasks as well as cytology image analysis \citep{zhang2017deeppap}.

Domain adaptation (DA) is another transfer learning approach that transfers knowledge by learning to narrow the distribution gap of datasets in different domains. The paradigm of DA can be defined as: giving two different datasets ($T$ and $S$) with different distributions ($p(S)\neq p(T)$), DA methods can align the distributions of these datasets by marginal, conditional or joint distribution adaptation. As a result, the knowledge is transferred from source to target domain, thus improving the performance of target models \citep{oza2021unsupervised}. In medical image analysis, data heterogeneity hinders the successful practice of deep models in the clinic. This issue can be mitigated by domain adaptation for improving the validity and reproducibility of deep models clinically \citep{guan2021domain}.

\begin{table*}[t!]
\scriptsize
\centering
\begin{center}
\caption{Summary of publicly available and representative private databases in computational cytology }
\label{Table_1} 
\begin{tabular}{p{2.75cm}p{1.5cm}p{1cm}p{0.75cm}p{1.5cm}p{3.5cm}p{4cm}}
\toprule

\multicolumn{1}{l}{Reference/Year}  & \multicolumn{1}{l}{Task} & \multicolumn{1}{l}{Organ} & \multicolumn{1}{l}{Stain} & \multicolumn{1}{l}{Size} &  \multicolumn{1}{l}{Description} &\multicolumn{1}{l}{Link}\\ \midrule 

\tabincell{l}{Herlev 2005\\ \citep{jantzen2005pap}} & Classification& Cervix &Pap& variable & 917 cells  & \url{http://mde-lab.aegean.gr/downloads}\\ \midrule 

\tabincell{l}{ISBI 2014\\ \citep{lu2015improved}} & Segmentation& Cervix &Pap&512 × 512 & 16 images (645 cells)  & \url{https://github.com/luzhi/cellsegmentation_TIP2015} \\ \midrule 

\tabincell{l}{ISBI 2015\\ \citep{lu2016evaluation}}  & Segmentation& Cervix &Pap&512 × 512 & 945 images synthesized by ISBI 2014  & \url{http://goo.gl/KcpLrQ} \\ \midrule 

\tabincell{l}{Sipakmed 2018\\ \citep{plissiti2018sipakmed}} &Classification & Cervix &Pap& 2,048 × 1,536&966 images (4,049 annotated cells) & \url{https://www.cs.uoi.gr/~marina/sipakmed.html}\\ \midrule 

\tabincell{l}{CERVIX93 2018\\ \citep{phoulady2018new}} &Classification Detection& Cervix &Pap& 1,280 × 960 &93 stacks of images (2,705 nuclei) & \url{https://github.com/parham-ap/cytology_dataset}\\ \midrule 

\tabincell{l}{FNAC 2019\\ \citep{saikia2019comparative}} &Classification& Breast &Pap& 2,048 × 1,536  &212 images in two classes: benign (99) and malignant (113)& \url{https://1drv.ms/u/s!Al-T6d-_ENf6axsEbvhbEc2gUFs} \\ \midrule 

\tabincell{l}{BHS 2019\\ \citep{araujo2019deep}} &Segmentation Ranking & Cervix &Pap& 1,392 × 1,040& 194 images in classes of carcinoma, HSIL, LSIL, ASC-US and ASC-H&\url{https://sites.google.com/view/centercric}\\ \midrule 

\tabincell{l}{AgNOR 2020\\ \citep{amorim2020novel}} & Segmentation& Cervix &AgNOR&1,600 × 1,200 & 2,540 images (4,515 nuclei)  & \url{https://arquivos.ufsc.br/d/373be2177a33426a9e6c/}\\ \midrule 

\tabincell{l}{LBC 2020\\ \citep{hussain2020liquid} } & Classification& Cervix &Pap&2,048 × 1,536 & 963 LBC images in classes of NILM, LSIL, HSIL, and SCC  &\url{https://data.mendeley.com/datasets/zddtpgzv63/4} \\ \midrule 

\tabincell{l}{Oral 2021\\ \citep{matias2021segmentation}} &Classification Detection Segmentation& Oral &Pap&1,200 × 1,600  &1,934 images (4,287 annotations) & \url{https://arquivos.ufsc.br/d/5035aec3c24f421a95d0/}\\ \midrule 

\tabincell{l}{Cric 2021\\ \citep{rezende2021cric}} & Classification& Cervix &Pap& 1,376 × 1,020& 400 images (11,534 cells)  & \url{https://database.cric.com.br} \\ \midrule 

\tabincell{l}{CDetector 2021\\ \citep{liang2021comparison}} & Detection& Cervix &Pap& 224 × 224& 7,410 images (48,587 object instance bounding boxes) in 11 classes  & \url{https://github.com/kuku-sichuan/ComparisonDetector} \\ \midrule 

\tabincell{l}{Ascites 2020\\ \citep{su2020development}} &Classification Detection& Stomach &H\&E, Pap&1,064 × 690  &487 images for classification in two classes: malignant (18,558) and benign (6,089). 176 images for detection (6,573 annotated cell bounding boxes)& \url{https://pan.baidu.com/s/1r0cd0PVm5DiUmaNozMSxgg}\\ \midrule 

\tabincell{l}{RSDC 2021\\ \citep{ma2021stsrnet}} &Super resolution & Cervix &Pap&\tabincell{l}{128 × 128 (HR) \\ 64 × 164 (LR)}& 5 slides (25000 patches) &  \url{https://www.kaggle.com/birkhoff007/rsdcdata}\\ \midrule 

\tabincell{l}{IRNet 2019\\ \citep{zhou2019irnet}} &Segmentation & Cervix &Pap&1,000 × 1,000&413 images (4,439 cytoplasm and 4,789 nuclei) & Private dataset\\ \midrule 

\tabincell{l}{DCCL 2020\\ \citep{zhang2019dccl}} &Detection & Cervix &Pap&1,200 × 2,000& 1,167 WSIs (14,432 patches, and 27,972 labeled lesion cells) & Private dataset\\ \midrule 

\tabincell{l}{Dual 2021\\ \citep{lin2021dual}} &Risk stratification & Cervix &Pap&Up to 50,000 × 50,000  & 19,303 WSIs in two classes: abnormal (202,557) and normal (272,933) & Private dataset\\ \midrule 

\tabincell{l}{Hybrid 2021\\ \citep{zhu2021hybrid}} &Classification Detection Segmentation & Cervix &Pap& 6000 × 6000 &24 categories and 2000 images in each category, 81,727 smears (1.7 million annotated targets) for detection model& Private dataset\\ 

\bottomrule
\end{tabular}
\end{center}
\end{table*}

\section{Datasets and metrics}\label{Datasets and metrics}
Deep learning relies on large amounts of labeled data, we summarize publicly available datasets as well as representative private datasets in cytology. As illustrated in Table \ref{Table_1}, the majority of public datasets are from the cervix, with a small number of other cancer types, such as breast, oral, and stomach. These publicly available cytology datasets can be used to develop deep learning algorithms for various tasks, including classification, detection, and segmentation.
\begin{table*}[h!]
\scriptsize
\centering
\begin{center}
\caption{Summary of evaluation metrics in computational cytology }
\label{Table_2} 
\begin{tabular}{p{2cm}p{1.8cm}p{5.8cm}p{6.8cm}}
\toprule
Metric & \multicolumn{1}{l}{Definition} & \multicolumn{1}{l}{Description}& \multicolumn{1}{l}{Application in cytology} \\ \midrule 
\textbf{Classification }&  & \\ \midrule

TP/TN/FP/FN &True Positive, True Negative, False Positive, False Negative & A test result that correctly indicates the presence of a condition (TP), correctly indicates the absence of a condition (TF), wrongly indicates the presence of a particular condition (FP), wrongly indicates the absence of a particular condition (FN). &Classification of FNAC images; formulate other metrics (e.g., accuracy, precision, recall) \citep{sanyal2018artificial,liang2021global}. \\ \midrule 

Confusion matrix &A Matrix. Row: actual class; Column: predicted class. &The number of correct and incorrect predictions are summarized with count values and broken down by each class.& Classification of lung cancer sub-type; pap smear image; quantitative analysis of abnormalities; WSI-level risk stratification \citep{teramoto2017automated, mohammed2021single,ke2021quantitative,awan2021deep}. \\ \midrule 

Accuracy (Acc)& $\frac{TP + TN }{FP + FN + TP + TN}$ &Proportion of all positive and negative classes with correct predictions in all samples.& Cervical squamous lesions classification; FNAC image classification \citep{liu2020artificial, bal2021bfcnet, albuquerque2021ordinal}.\\ \midrule 

Precision (P)& $\frac{TP}{FP + TP} $ &The proportion of positive samples classified as positive examples by the classifier.&
Cervical cells classification; cervical lesions classification; multi-cell classification in liquid-based cytology images \citep{sornapudi2019comparing,liu2020artificial,rahaman2021deepcervix}. \\ \midrule 

Recall (R)& $\frac{TP}{FN + TP} $ &The proportion of the samples predicted to be positive cases in the total positive cases.& Pap smear image classification; classification of cervical cells \citep{mohammed2021single,rahaman2021deepcervix}.  \\ \midrule 

Specificity (Spec)& $\frac{TN}{FP + TN} $ &The proportion of samples that are correctly predicted as negative classes in all negative classes.& Cell classification; differential diagnosing of papillary thyroid carcinomas; detection of cervical intraepithelial neoplasia or invasive cancer \citep{zhang2017deeppap,guan2019deepb,bao2020artificial}. \\ \midrule 

Sensitivity (Sens)& $\frac{TP}{FN + TP} $&The proportion of the samples predicted to be positive cases in the total positive cases.& Distinguish large cell neuroendocrine; identify cells; high resolution image classification \citep{gonzalez2020feasibility,li2021mixed}. \\ \midrule 

F1-score (F1)& $\frac{ 2 \times Precision \times Recall}{Precision + Recall}$  =$\frac{2 \times TP}{FP+FN+2 TP}$&Harmonic average of precision and recall, and it is defined as the final evaluation index in many classification tasks.& Classification of cervical cells; multi-cell classification in liquid-based cytology images; cell image ranking \citep{sornapudi2019comparing,araujo2019deep,rahaman2021deepcervix}. \\ \midrule 

ROC curve& (FP rate, TP rate)&ROC is a graph showing the performance of a classification model at all classification thresholds. & Prediction of malignancy; cervical cancer screening; smear-level risk stratification \citep{elliott2020application, tan2021automatic,lin2021dual}.  \\ \midrule 

AUC& Area under the ROC curve& The closer the AUC is to 1, the better the classifier performance.&Cancer screening (cell-level detection, patch-level and case-level classification); quantitative analysis of abnormalities; automating the paris system for cytopathology \citep{vaickus2019automating,ke2021quantitative,cao2021novel}. \\ \midrule 

\textbf{Detection} &  & \\ \midrule 
IoU & $\frac{P\cap GT}{P\cup GT}$  $= \frac{TP}{FP+FN+TP}$ &$P$ denotes predicted bounding box, and GT is ground truth box. The ratio of the intersection and union of the predicted bounding box and the ground truth bounding box.&Nuclei/Cell detection; automation-assisted cervical cancer reading \citep{ kilic2019automated,xiang2020novel,liang2021global}. \\ \midrule 

AP  & Average precision & The mean value of precision on precision-recall curve. &Detection of abnormal cervical cells \citep{cao2021novel}. \\ \midrule 

mAP& mean AP & Average of AP in all categories. & Cell/Clumps detection; quantification of pulmonary hemosiderophages \citep{marzahl2020deep,chai2021deep,liang2021comparison}. \\ \midrule

\textbf{Segmentation}&  & \\ \midrule 
Pixel Precision (P$_{p}$)& $\frac{T P_{p}}{T P_{p}+F P_{p}}$& The ${p}$ means this is a pixel-level metric. Proportion of correctly segmented pixels to all segmented pixels.&Cytological examination (overlapping cell segmentation) \citep{tareef2017optimizing}. \\ \midrule
 
Pixel Recall (R$_{p}$) & $\frac{T P_{p}}{T P_{p}+F N_{p}}$ &Proportion of correctly segmented pixels to all pixels in the ground truth.&Cytological examination (overlapping cell segmentation) \citep{tareef2017optimizing}. \\ \midrule 

Pixel Accuracy (Acc$_{p}$) &$ \frac{TP_{p} + TN_{p} }{FP_{p} + FN_{p} + TP_{p} + TN_{p}}$ &Pixel level accuracy.&Segmentation of cytoplasm and nuclei \citep{song2015accurate,song2014deep,ke2021quantitative}. \\ \midrule 

Hausdorff Distance & $\max ( \sup \limits_{x \in X} d(x, Y)$, $\sup \limits_{y \in Y} d(X, y) )$ & X and Y are two sets,  $sup$ represents the supremum. It measures the similarity between two point sets.&Cell nuclei segmentation \citep{kowal2020cell}.  \\ \midrule 

Dice coefficient (Dice) & $\frac{2 \times TP}{FP+FN+2 TP}$ &Dice coefficient is a statistical tool which measures the similarity between two sets of data. It can be used for comparing algorithm output against reference masks.&Semantic instance segmentation of touching and overlapping objects; cytoplasm segmentation; instance segmentation \citep{bohm2019isoo,wan2019accurate,walter2021multistar}. \\  \midrule 

Zijdenbos similarity index (ZSI) & $2 \frac{\left|R_{G T} \cap R_{S e g}\right|}{\left|R_{G T}\right|+\left|R_{S e g}\right|}$& $R_{G T} $ and $R_{S e g}$ denote the ground truth and segmented regions, respectively. ZSI computes the ratio of aggregated union between cardinality predicted segmentation output and manual segmentation output. & Overlapping cell segmentation; cervical nuclei segmentation \citep{tareef2017optimizing,hussain2020shape}.\\ \midrule 

Average Jaccard Index (AJI)  & $\frac{\sum_{i=1}^{N}\left|G_{i} \cap P_{M}^{i}\right|}{\sum_{i=1}^{N}\left|G_{i} \cup P_{M}^{i}\right|+\sum_{F \in U}\left|P_{F}\right|}$ & $G_{i} $ is the $i\text{-}{th}$ object from the ground truth with $N$ objects. $P_{M}^{i}$ means the $M\text{-}{th}$ connected component in prediction which has the largest Jaccard Index with $G_{i} $. AJI measures the ratio of the aggregated intersection and aggregated union for all the predictions and ground truths in the image.&Cell segmentation \citep{zhou2019irnet}. \\

\bottomrule
\end{tabular}
\end{center}
\end{table*}

\hangafter 1
\hangindent 1em
\noindent
\textbf{Herlev} \citep{jantzen2005pap}. This database consists of 917 Papanicolaou (Pap) smear cervical images in 7 classes (3 normal cell classes and 4 abnormal cell classes), which are collected from the Herlev University Hospital. As the earliest established public cytology dataset, Herlev dataset is extensively adopted for developing DL-based coarse and fine-grained classification models for cervical cancer screening \citep{zhang2017deeppap,lin2019fine}.

\hangafter 1
\hangindent 1em
\noindent
\textbf{ISBI 2014} \citep{lu2015improved}. Another widely developed cervical dataset comes from the ISBI challenge. Different from Herlev dataset, this dataset focuses on the segmentation task with pixel-wise annotations. It consists of 16 non-overlapping fields of view images (×40 magnification) with 645 cells obtained from four cervical cytology specimens. Each sample in this dataset contains 20 to 60 Pap-stained cervical cells with varying degrees of overlapping. 

\hangafter 1
\hangindent 1em
\noindent
\textbf{ISBI 2015} \citep{lu2016evaluation}. ISBI 2015 extends ISBI 2014 dataset to 945 cervical cytology images by synthesizing. ISBI 2015 has a varying number of cells and different degrees of cell overlapping (size of 512 × 512 pixels), which contains 45 training images (taken from the 4 extended depth field images) and 900 testing images (from 12 images). 

\hangafter 1
\hangindent 1em
\noindent
\textbf{Sipakmed} \citep{plissiti2018sipakmed}. This database consists of 4049 images of isolated cells that have been manually cropped from 966 cluster cell images of Pap smear slides. Sipakmed dataset has 5 types of cervical cells, including superficial-intermediate, parabasal, koilocytotic, dyskeratotic, and metaplastic cells.

\hangafter 1
\hangindent 1em
\noindent
\textbf{CERVIX93} \citep{phoulady2018new}. This is the first dataset established for nuclei detection tasks in cytology. It consists of 93 stacks of images at 40$\times$ magnification. Each stack has 10-20 images acquired at the equally spaced field of views from the top to the bottom of the slide. In this dataset, 2705 nuclei are annotated by bounding boxes with three different Pap test grades: negative, low-grade squamous in the intraepithelial lesion (LSIL) or high-grade squamous intraepithelial lesion (HSIL).

\hangafter 1
\hangindent 1em
\noindent
\textbf{FNAC} \citep{saikia2019comparative}. This is the only public breast cytology dataset developed for classification model. These breast images are collected from 20 patients, comprising of 212 fine-needle aspiration cell inspection images in classes of benign (99) and malignant (113).

\hangafter 1
\hangindent 1em
\noindent
\textbf{BHS} \citep{araujo2019deep}. It collects 194 Pap-smear cervical slides from the Brazilian Health System (BHS). Among them, 108 images have at least one abnormal cell and 86 images with normal cells only. In sum, it has 5 types of abnormalities: carcinoma, HSIL, LSIL, atypical squamous cells of undetermined significance (ASC-US) and atypical squamous cells cannot exclude HSIL (ASC-H).

\hangafter 1
\hangindent 1em
\noindent
\textbf{AgNOR} \citep{amorim2020novel}. The dataset is composed of 2540 images with $1200 \times 1600$ pixels each. Different from other public cervical datasets, it contains cells stained with the silver technique, which is known as argyrophilic nucleolar organizer regions (AgNOR). For developing segmentation approaches, experts annotate objects by the Labelme tool \citep{russell2008labelme}, including nuclei, clusters, and satellites.

\hangafter 1
\hangindent 1em
\noindent
\textbf{LBC} \citep{hussain2020liquid}. Recently, liquid-based cytology (LBC) is developed for providing more uniform fixation with a cleaner background and well-preserved samples than conventional Pap smear tests. This dataset consists of 963 images with four classes: NILM, LSIL, HSIL, and squamous cell carcinoma (SCC).

\hangafter 1
\hangindent 1em
\noindent
\textbf{Oral} \citep{matias2021segmentation}. Totally, 1,934 oral images of $1200 \times 1600$ pixels are acquired from two Pap-stained slides of cancer diagnosed oral brush samples. With different types of annotation (category, box, mask), various DL tasks can be conducted by this dataset, including classification, detection, and segmentation. 

\hangafter 1
\hangindent 1em
\noindent
\textbf{CRIC} \citep{rezende2021cric}. The collection CRIC cervix has 400 images of pap smears with 11,534 classified cells. Based on the Bethesda system \citep{nayar2015bethesda}, CRIC dataset covers conventional cytology cervical cells with six types: NILM (6,779), ASC-US (606), LSIL (1,360), ASC-H (925), HSIL (1,703), and SCC (161).

\hangafter 1
\hangindent 1em
\noindent
\textbf{CDetector} \citep{liang2021comparison}. This dataset consists of 7,410 cervical images cropped from the WSIs. According to the Bethesda system (TBS), 48,587 object instance bounding boxes are annotated by experienced pathologists which belong to 11 categories: ASC-US, ASC-H, HSIL, LSIL, SCC, atypical glandular cells (AGC), trichomonas (TRICH), candida (CAND), flora, herps, actinomyces (ACTIN). Till now, CDetector is the largest public dataset for the object detection task in cytology.

\hangafter 1
\hangindent 1em
\noindent
\textbf{Ascites} \citep{su2020development}. This dataset is established for screening gastric cancer and collected from Peking University. It consists of 176 H\&E stained and Pap stained images cropped from ascites cytopathology images at 40 × magnification. A total of 6573 cells (benign and malignant) are annotated using bounding boxes.

\hangafter 1
\hangindent 1em
\noindent
\textbf{RSDC} \citep{ma2021stsrnet}. It is the only public cytology dataset for developing the refocusing and super-resolution task. The images in the dataset are collected from 5 LBC slides with the resolution 0.243$ \mu m/pixel$. Strategies of bicubic interpolation and gaussian blur are used to generate 15,000 low-resolution images (64 × 64) and corresponding high-resolution images (128 × 128) from original slides.

Furthermore, to evaluate the performance of proposed deep learning models, we summarize the evaluation metrics in terms of three canonical DL approaches: classification, detection, and segmentation along with typical cytological applications adopting these metrics, more details are shown in Table \ref{Table_2}.

Among these summarized evaluation metrics, classification metrics evaluate the classifier's capability of predicting the category. Accuracy is the most straightforward metric, yet it ignores the imbalance problem between different categories. The confusion matrix can represent the prediction result of each category. To measure the detection task, $IoU$ is a commonly-used metric, which can measure the overlap between the predicted box and the ground truth box. Based on various set thresholds, average precision ($AP$) is utilized to evaluate the performance of the detector in different overlapping levels, including $AP_{50}$, $AP_{75}$, etc. Segmentation models can be evaluated by various metrics. For example, $Pixel\ accuracy$ measures the predicted result of each pixel, and $Dice$ calculates the similarity coefficient of predicted masks and ground truth.
\section{Deep learning in cytology application}\label{Deep learning in cytology application}
In this section, we survey and summarize literatures on various deep learning models applied in computational cytology. Firstly, we introduce preprocessing techniques in cytology image analysis, followed by representative clinical tasks: classification, detection, segmentation, and others. More details of these surveyed literatures are as follows.

\subsection{Preprocessing}\label{Preprocessing}

\textbf{Staining techniques.} In cytology, staining techniques are introduced to enhance the image features of cells (e.g., texture, structure, and biochemical properties) for visually presenting cellular structures. 

\begin{figure}
		\centering
		\includegraphics[width=0.49\textwidth]{{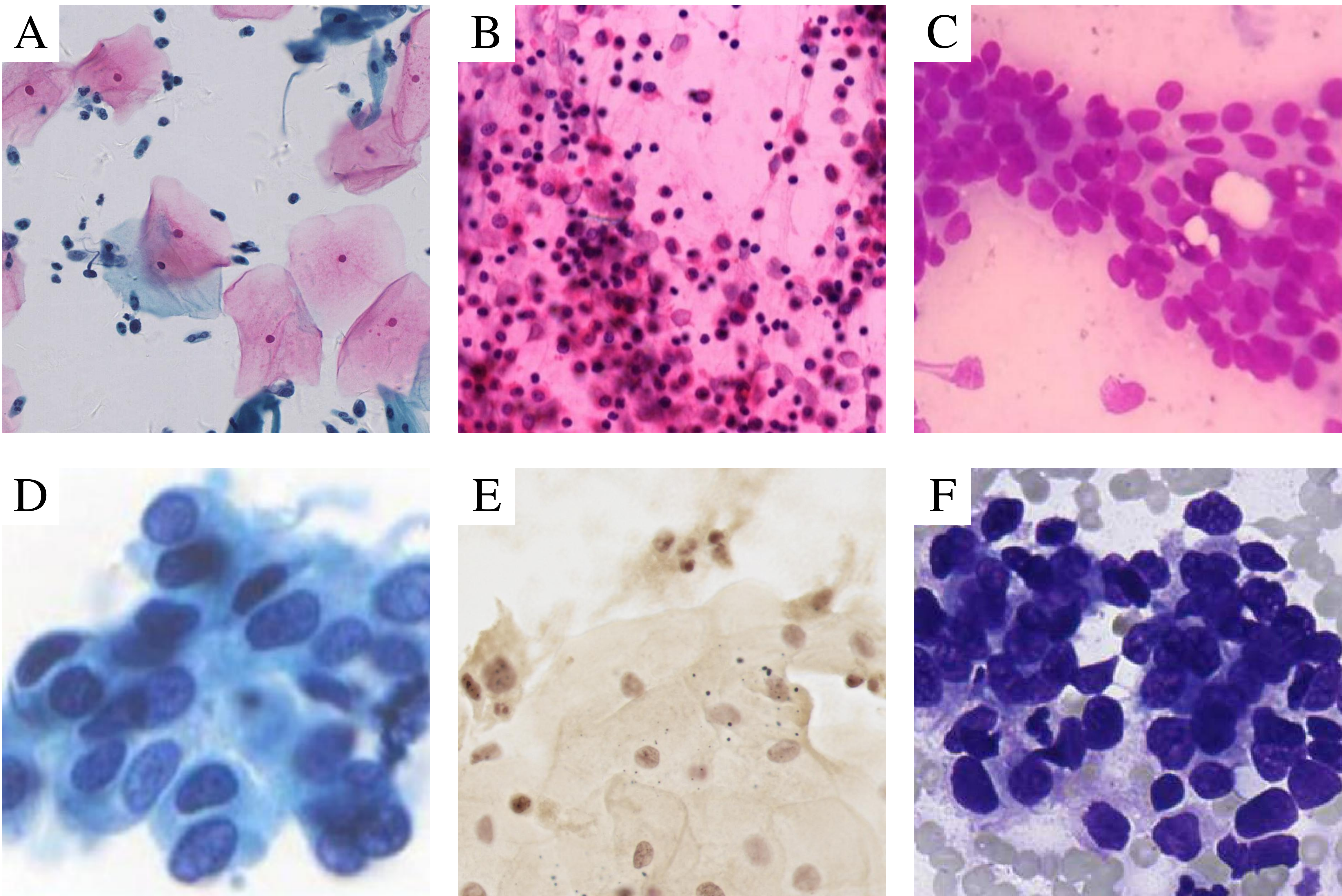}}
		\caption{Staining techniques. (A) Pap \citep{zhou2019irnet}. (B) H\&E \citep{su2020development}. (C) Giemsa \citep{bal2021bfcnet}. (D) Feulgen \citep{xiang2020novel}. (E) AgNOR \citep{amorim2020novel}. (F) Diff-quik \citep{gonzalez2020feasibility}.}
		\label{Fig_8}
	\end{figure}

Fig. \ref{Fig_8} shows various staining techniques in cytology. 1) Pap. As the extensive staining protocols, it has four steps: fixation, nuclear staining, cytosol staining, and transparency. Cell stained by Pap has a clearly-structured nucleus and transparent cytoplasm. According to surveyed literatures, Pap is the most common staining method in cytology images, especially in cervical cancer. 2) Hematoxylin and Eosin (H\&E). Hematoxylin stains cell nuclei purplish-blue, and eosin stains the extracellular matrix and cytoplasm pink. In the clinic, H\&E is mainly used for staining cells and tissues. 3) Giemsa. It is particularly effective for staining cytoplasm, so Giemsa is mainly used for blood and bone marrow cytological evaluation. Other staining techniques are used for some specific situations. For example, \cite{amorim2020novel} stained cervical cells by AgNOR to present cell proliferation, differentiation, and malignant transformation. In another work, \cite{xiang2020novel} stained cervical specimens of LBC by Feulgen. \cite{marzahl2020deep} stained pulmonary hemosiderophages cytology by Perlss’ Prussian Blue, Turnbull’s Blue. 
	
\textbf{Stain Normalization.} A significant amount of color variations exist in cytology images due to various staining techniques mentioned and other issues (e.g., imaging environment). These differences bring challenges for building robust and generated DL-based cytology models. Besides, normalization can accelerate the convergence when training networks. Therefore, normalization can be a crucial preprocessing step, especially when analyzing stained images, like cytology images, and histopathology images. Commonly-used methods include whiting for removing redundant information from input data, and linear normalization for scaling gray values of input data, etc.

\textbf{Data augmentation.} When the amount of images is not sufficient to learn a robust DL model, especially for medical images. Data augmentation strategies are introduced to increase the amount of input images for improving the model's generalization. Conventional augmentation methods include geometric transformation (e.g., flipping, rotating, and scaling) and color transformation (e.g., noise, blurring, and contrast). Recently, generative adversarial network (GAN) has been adopted to synthesize a large-scale dataset based on a limited set. \cite{yu2021generative} utilized GAN to generate 16,000 images from 961 real images for improving cervical cell classification models. In another study, \cite{dey2019syncgan} generated 180 images by conditional GAN and learnable class-specific priors for improving classifier performance on cytology tasks. To overcome the data limitation issue, \cite{teramoto2020deep} proposed a GAN-based augmentation structure, progressive growing of GANs (PGGAN). In this study, real lung cytological images together with synthesized images by PGGAN are used to train classification CNNs, leading to the performance improvements in cytology image classification.

\subsection{Classification}
In cytology, the DL approaches are feasible and promising for image classification with distinguishable characteristics of cytology samples. The clinical cytologists can distinguish between benign and malignant cells based on their cytological features. For example, the abnormalities are displayed in malignant cells, such as larger and irregular nuclei, enlarged nuclear-cytoplasmic ratio, and the altered shape of the nucleolus. Within cytology image classification tasks, DL-based models aim to extract the underlying patterns of input images for identifying objects (e.g., nucleus, cell) or making slide-level predictions (e.g., cytopathology screening). Therefore, we further divide the cytological classification task into two categories: 1) cell-level, 2) slide-level.

\begin{table*}[htbp]
\scriptsize
\centering
\begin{center}
\caption{Overview of deep learning-based classification studies for computational cytology}
\label{Table_3} 
\begin{tabular}{p{1.5cm}<{\centering}p{2.5cm}p{0.75cm}p{0.75cm}p{3cm}p{3cm}p{3.5cm}<{\raggedright}}
\toprule
Reference & \multicolumn{1}{l}{Application} & \multicolumn{1}{l}{Staining} & \multicolumn{1}{l}{Organ} & \multicolumn{1}{l}{Method} & \multicolumn{1}{l}{Dataset}& \multicolumn{1}{l}{Result}\\ \midrule 

\multicolumn{2}{l}{\textbf{Cell-level classification} }&&&&& \\ \midrule 
\cite{zhang2017deeppap}&Cell classification&Pap H\&E&Cervix&Data preprocessing (patch extraction, data augmentation) + CNN + Transfer learning (fine-tune)&Herlev; HEMLBC	&Herlev: Sens=0.982, Spec=0.983, Acc=0.983, F1=0.988, AUC=0.998;  \qquad  \qquad \qquad HEMLBC: Sens=0.983, Spec=0.990, Acc=0.986.\\ \midrule 

\cite{teramoto2017automated}&	Classification of cancer types (adenocarcinoma, squamous cell carcinoma, and small cell carcinoma)&	Pap&	Lung	&Data augmentation + CNN	&Private dataset: 76 images in classes of adenocarcinoma (40), squamous cell carcinoma (20), and small cell carcinoma (16)&Adenocarcinoma: Acc=0.89; Squamous cell carcinoma: Acc=0.600;	Small cell carcinoma: Acc=0.703; Total: Acc=0.711. \\ \midrule

\cite{dimauro2019nasal}&	Cell classification&Pap&Nasal&Three-block CNN 	&Private dataset: 3,423 images (cell)	&Sens=0.97, Acc=0.99.	 \\ \midrule

\cite{shanthi2019deep}&Malignancy detection and classification&Pap&Cervix&Three-layer CNN&Herlev&5-class: Acc=0.941; 4-class: Acc=0.962; 3-class: Acc=0.948; 2-class: Acc=0.957.\\ \midrule

\cite{teramoto2019automated}&	Classification of benign and malignant cells &Pap&Lung&Data augmentation + VGG-16 + GradCAM&Private dataset: 621 images (patch) in classes of benign (306) and malignant (315)&Patch-level: Acc=0.792, AUC=0.872; Case-level: Acc=0.870, AUC=0.932. \\ \midrule

\cite{liu2020artificial}&Classification of cervical squamous lesions&Pap&Cervix&VGG-16&Private dataset: 3,290 images in classes of abnormal cells (1,736 ) and normal cells (1,554)&Acc=0.9807, P=0.9791, Sens=0.9801, F1=0.9809.\\ \midrule

\cite{lin2019fine}&Fine-grained cell classification&Pap&Cervix&Fine-tune + CNN (AlexNet, GoogLeNet, ResNet, and DenseNet)&Herlev&GoogLeNet: Acc=0.945 (2-class), Acc=0.713 (4-class), Acc=0.645 (7-class).	\\ \midrule 

\cite{sornapudi2019comparing}&Multi-cell classification in liquid-based cytology images&Pap&Cervix&Fine-tune + CNN (ResNet-50, VGG-19, DenseNet-121,  Inception-v3)&Herlev; Private dataset: 25 images&ResNet-50: F1=0.8865, AUC=0.95; VGG-19: F1=0.8896, AUC=0.95; Densenet-121: F1=0.8546, AUC=0.94; Inception-v3: F1=0.8072, AUC=0.88. \\ \midrule

\cite{hussain2020comprehensive}&Cervical cancer diagnostic prediction&Pap&Cervix&CNN (AlexNet, VGG-16, VGG-19, ResNet-50, ResNet-101, and GoogLeNet)&Herlev&AlexNet: Acc=0.8; \qquad \qquad VGG-16: Acc=0.8337; \qquad VGG-19: Acc=0.8455; \qquad ResNet-50: Acc=0.8937; ResNet-101: Acc=0.9450; GoogLeNet: Acc=0.9567.\\ \midrule

\cite{mohammed2021single}&Smear classification&Pap&Cervix&10 popular pre-trained 
CNN &Sipakmed&DenseNet-169: Acc=0.990, P=0.974, R=0.974, F1=0.974.\\ \midrule

\cite{albuquerque2021ordinal}&Classification of cervical cancer risk&Pap&Cervix&9 popular CNN&Herlev&Acc=0.756 (7-class), Acc=0.813 (4-class).\\ \midrule

\cite{miselis2019deep}&	Classification of FANC images&H\&E&Breast&CNN (AlexNet, GoogLeNet, SqueezeNet, DenseNet, Inception-V3)&Private dataset: 737 images (ROIs from specimens) in classes of benign (275) and malignant (462)&AlexNet: AUC=0.9730; GoogLeNet: AUC=0.9455; SqueezeNe: AUC=0.9152; DenseNet: AUC=0.9244; Inception-V3: AUC=0.9730.	\\ \midrule

\cite{manna2021fuzzy}&Classification of cervical cells&Pap&Cervix&Fuzzy rank + Pre-trained CNN (Inception-V3, Xception and DenseNet‑169)&Sipakmed&Acc=0.9855, Sens=0.9852.\\ \midrule

\cite{rahaman2021deepcervix}& Classification of cervical cells&Pap&Cervix&Hybrid deep feature fusion + CNN (VGG-16, VGG-19, XceptionNet, and ResNet-50)&Sipakmed&Acc=0.9985 (2-class), Acc=0.9914 (3-class), Acc=0.9914 (5-class).\\ \midrule

\cite{yu2021generative}&Classification of cervical cells&Pap&Cervix&AlexNet + GAN&Private dataset: 22,124 images (cell) in classes of abnormal (1,202) and normal (20,922)&AUC=0.984\\ \midrule

\cite{dey2019syncgan}&FNAC cytology image classification&H\&E&Breast&Conditional GAN (synthesis) + CNN (ResNet-152, DenseNet-161, Inception-V3)&Private dataset: 150 images in classes of begin (75) and malignant (75)&180 generated images. ResNet-152: Acc=0.7667; DenseNet-161: Acc=0.8667; Inception-V3: Acc=0.8000.\\ \midrule

\cite{teramoto2020deep}&Classification of cytological images&	Pap&Lung&CNN + PGGAN&Private dataset: 511 images (patch) in classes of benign (244) and malignant (267)&Acc=0.853, Sens=0.854, Spec=0.853. \\ \midrule

\cite{bakht2020thyroid}&Classification of FNAC images&Pap&Thyroid&CNN (VGG-19, AlexNet) + Transfer learning (Fine-tune)&Private dataset:  9,209 images (cell) in 5 classes&VGG-19: Acc=0.9305; AlexNet: Acc=0.9288. \\ \midrule

\cite{vaickus2019automating}&Automating the Paris system for cytopathology&Pap&Urine&VGG-19 + Morphometric model&Private dataset: 217 WSIs in classes of negative (51), atypical (60), suspicious (52), and positive (54)&Acc=0.972, Spec=0.976, Sens=0.970.\\ 
\bottomrule

\multicolumn{7}{r}{\footnotesize\textit{continued on the next page}}\\
\end{tabular}
\end{center}
\end{table*}

\begin{table*}[htbp]
\ContinuedFloat
\centering
\scriptsize
\begin{center}
\caption{Overview of deep learning-based classification studies for computational cytology (continued)}
\begin{tabular}{p{1.5cm}p{2.5cm}p{0.75cm}p{0.75cm}p{3cm}p{3cm}p{3.5cm}<{\raggedright}}
\toprule

Reference & \multicolumn{1}{l}{Application} & \multicolumn{1}{l}{Staining} & \multicolumn{1}{l}{Organ} & \multicolumn{1}{l}{Method} & \multicolumn{1}{l}{Dataset}& \multicolumn{1}{l}{Result}\\ \midrule 

\cite{kaneko2021urine}&Cell image recognition&Pap&Urine&EfficientNet&Private dataset: 4,637 images (cell)&Acc=0.95, Sens=0.97, Spec=0.95, and AUC=0.99.\\ \midrule

\cite{zejmo2017classification}&Classification of cancer cytological specimen&H\&E&Breast&CNN (AlexNet, GoogLeNet)&550 images (ROIs) in classes of malignant (275) and benign (275)&AlexNet: Acc=0.80; GoogLeNet: Acc=0.83.\\ \midrule

\cite{shi2021cervical}&Classification of cervical cells&Pap&Cervix&Graph convolutional Network (GCN)&Sipakmed&Acc=98.37 ± 0.57, Sens=99.80 ± 0.10, Spec=99.60 ± 0.20, F1=99.80 ± 0.10.\\ \midrule

\cite{garud2017high}&Classification of FANC cell samples&H\&E&Breast&GoogLeNet& Private dataset: 37 images in classes of benign (24) and malignant (13)&Acc=0.8076.	 \\ \midrule

\cite{saikia2019comparative}&Classification of FNAC images&Pap&	Breast&CNN (VGG-16, VGG-19, ResNet-50, and GoogLeNet-V3)& FANC 2019&VGG-16: Acc=0.8867; VGG-19: Acc=0.882; ResNet-50: Acc=0.9056;  GoogLeNet-V3: Acc=0.9625. 	\\ \midrule

\cite{nojima2021deep}&Diagnose the malignant potential of carcinoma cells&Pap&Urine&Visual geometry group CNN&Private dataset: 203 images&AUC=0.9890, F1=0.9002.\\ \midrule

\cite{lilli2021calibrated}& Cell classification&Pap&Urine&VGG-16&Private dataset: 690 images in classes of urothelial normal cells (274) and abnormal cells (416)&Acc=0.899.	 \\ \midrule

\cite{guan2019deepb}&Differential diagnosing of papillary thyroid carcinomas&H\&E&Thyroid&VGG-16 and Inception-v3&Private dataset: 279 images (thyroid nodules)&VGG-16: 0.9766 (image-level), 0.95 (patient-level); Inception-v3: 0.9275 (image-level), 0.875 (patient-level).\\ \midrule

\cite{bal2021bfcnet}&Classification of FNAC images&Giemsa H\&E&Breast&CNN (13 layers, convolution and fully-connected layers)&Private dataset: Giemsa (1020 images in classes of benign and malignant) and H\&E (631 images in classes of benign and malignant) &Giemsa: Acc=0.9781, P=0.977, R=0.973, Spec=0.982, F1=0.975; H\&E: Acc=0.9753, P=0.973, R=0.950, Spec=0.987, F1=0.961.	 \\ \midrule

\cite{guan2019deep}&Differential diagnosing of lymph node&H\&E&Cervix&Inception-v3&Private dataset: 742 images in 4 classes&Acc=0.8962.\\ \midrule

\cite{bhatt2021cervical}&Cervical cancer detection&Pap&Cervix&EfficientNet + Grad-CAM&Herlev, Sipakmed&Acc=0.9970, P=0.9970, R=0.9972, F1=0.9963, Kappa=0.9931.\\ \midrule

\cite{noyan2020tzancknet}&Cell identification&Giemsa&Skin&ResNet-50&Private dataset: 2,260 images (Tzanck smear)&Acc=0.943, Sens=0.837, Spec=0.973, AUC=0.974.\\ \midrule

\cite{bao2020artificial}&Detection of cervical intraepithelial neoplasia or invasive cancer&Pap&Cervix&VGG-16&Private dataset: 188,542 mages&CIN 2: Acc=0.926; CIN 3+: Acc=0.961.\\  \midrule

\cite{sanyal2018artificial}&Classification of FNAC images&Giemsa Pap&Thyroid&CNN&Private dataset: 370 images in classes of non‑PTCA (184) and PTCA (186)	&Sens=0.9048, Spec=0.8333, Acc=0.8506. \\ \midrule

\cite{wu2018automatic1}&Classification of cancer types&H\&E&Cervix&AlexNet&Private dataset: 79 specimens in 3 classes&Acc=0.9333.\\ \midrule

\cite{li2022cervical}&Cervical cell classification&Pap&Cervix&ResNet-50 + Attention mechanism + LSTM&Sipakmed&Sensitivity=0.999, specificity=0.998, F1=0.9989. \\  \midrule

\multicolumn{2}{l}{\textbf{Slide-level classification} }&&&&& \\ \midrule 

\cite{elliott2020application}&Classification (prediction of malignancy)&Pap&Thyroid&AlexNet	&Private dataset: 908 WSIs &Sens=0.92, Spec=0.905, AUC=0.932.	 \\ \midrule

\cite{dov2019thyroid}&Classification (prediction of malignancy)&Pap&Thyroid&VGG-11 + Multiple instance learning &Private dataset: 908 WSIs&AUC=0.932, AP=0.872. \\ \midrule

\cite{tan2021automatic}&Cervical cancer screening&Pap&Cervix&Faster R-CNN&Private dataset: 408,030 images&Sens=0.994, Spec=0.348, AUC=0.67.\\ \midrule

\cite{ke2021quantitative}&Quantitative analysis of abnormalities&Pap&Cervix&U-Net, ResNet-50&Private dataset: 130 WSIs &Segmentation: Pixel Acc=0.974 ± 0.001, IoU=0.913 ± 0.007; Classification: Acc=0.945 ± 0.006.\\ \midrule

 \cite{cao2021novel}&Cancer screening (cell-level detection, patch-level and case-level classification)&Pap&Cervix&Multi-scale region-based CNN + Attention mechanism&Private dataset: 7030 images&Cell-level detection: AP=0.7509; Patch-level classification: AUC=0.9909; Case-level classification: AUC=0.934, Sens=0.913, Spec=0.906 and Acc=0.909.\\ \midrule

\cite{li2021mixed}&High resolution image classification&Pap&Cervix&Mixed supervision learning (image-level + pixel-level)& Private dataset: 862 images from 2 data centers&Center A: Sens=1, Spec=0.86; Center B: Sens=1, Spec=0.87. \\ 

\bottomrule

\multicolumn{7}{r}{\footnotesize\textit{continued on the next page}}\\
\end{tabular}
\end{center}
\end{table*}

\begin{table*}[htbp]
\ContinuedFloat
\centering
\scriptsize
\begin{center}
\caption{Overview of deep learning-based classification studies for computational cytology (continued)}
\begin{tabular}{p{1.5cm}p{2.5cm}p{0.75cm}p{0.75cm}p{3cm}p{3cm}p{3.5cm}<{\raggedright}}
\toprule

Reference & \multicolumn{1}{l}{Application} & \multicolumn{1}{l}{Staining} & \multicolumn{1}{l}{Organ} & \multicolumn{1}{l}{Method} & \multicolumn{1}{l}{Dataset}& \multicolumn{1}{l}{Result}\\ \midrule 

\cite{dov2021weakly}&Classification (prediction of malignancy)&Pap	&	Thyroid&	MIL+ NoisyAND + Attention mechanism + Maximum likelihood estimation&Private dataset:  142 WSIs with 4,494 instances&AUC=0.870 ± 0.017, AP=0.743 ± 0.037.	\\ \midrule

\cite{gonzalez2020feasibility}&Distinguish large cell neuroendocrine&Pap, H\&E, Diff-Quik&Lung&CNN&Private dataset: 40 images in high-grade neuroendocrine carcinoma (17 small cell, 13 large cell, 10 mixed/unclassifiable)&H\&E: Acc=0.900; Pap: Acc=0.875; Diff-Quik: Acc=0.889.\\ \midrule

\cite{zhu2021hybrid}&Rapid TBS classification of cervical liquid-based thin-layer cell smears&Pap&Cervix&Model assembly: Xception (classification), YOLOv3 (object detection), and U-Net (segmentation)&Private dataset: 81,727 images& Speed=180s/slide, Sens=0.9474. \\ \midrule

\cite{wei2021efficient}& Cervical lesion detection, WSI-level classification of normal and abnormal &Pap&Cervix&YOLOv3 + Transformer&Private dataset: 2,019 images (slide) from four scanning devices&AUC=0.872. \\ \midrule	

\cite{cheng2021robust}& WSI-level cervical cancer screening &Pap&Cervix&CNN (ResNet50) + RNN&Private dataset: 3,545 images (slide) with 79,911 annotations&Spec=0.935, Sens=0.951, Speed=1.5min/slide. \\ \midrule	

\cite{awan2021deep}&	Cell-level classification, WSI-level risk stratification&Pap&	Urine&RetinaNet + Counting of atypical and malignant cells&Private dataset: 398 images (slide)  in classes of normal (243), inflammatory (13), CA (76), ASM (38) and TCC (28)&Cell-level classification: AUC=0.99; Risk stratification: AUC=0.83.\\ \midrule	

\cite{lin2021dual}&Smear-level risk stratification&Pap&Cervix&CNN with dual-path encode + Synergistic grouping loss&Private dataset: 19,303 WSIs (13,486 for training, 2,486 for validation and 3,331 for testing) from 4 centers in 6 classes of cells& Sens=0.907, 
Spec=0.80, AUC=0.925.\\ 

\bottomrule
\end{tabular}
\end{center}
\end{table*}

\subsubsection{Cell-level classification}
Cell-level classification could be one of the most successful tasks in DL-based cytology image analysis \citep{jantzen2005pap}. Due to the giga-pixel resolution of collected cytology WSIs, they are usually cut into cell patches for image analysis \citep{zhang2019dccl}. When training a deep network for classifying cells, cell patches are cropped from these whole images first. Then, these cell patches are fed into DL models to train cell-level classification models after preprocessing.

The most straightforward method is to directly feed cell patches into a multi-layer CNN for extracting feature maps, then crossing the output layer to get the predicted category. A series of CNN-based methods have been proposed. For lung cytology classification, \cite{teramoto2017automated} designed a deep convolutional neural network consisting of three convolutional layers, three pooling layers, and two fully connected layers. Similarly, \cite{dimauro2019nasal} constructed a three-block CNN model for nasal cell classification. For cervical cytology, \cite{shanthi2019deep} designed a CNN architecture consisting of three convolutional layers. Its experimental results in different settings (2 class, 3 class, 4 class, and 5 class) showed an effective performance of different grades of cancer in cervical images. In addition, \cite{zhang2017deeppap} proposed a simple ConvNet, which was first pre-trained in a natural image dataset, ImageNet. Then, the model was fine-tuned in two cervical cytological datasets, Herlev and HEMLBC \citep{jantzen2005pap,zhang2014automation}, achieving outperforming performance than previous algorithms. However, the performances and generalization capabilities of these simply-designed CNN with several layers are limited to specific datasets and scenarios.

A large amount of advanced deep models are proposed in the computer vision field, such as Inception \citep{szegedy2015going}, ResNet \citep{he2016deep} and DenseNet \citep{huang2017densely}. These networks can be directly adopted for cytology image analysis and achieve better performance than simply-designed structures in the classification task. For example, \cite{teramoto2019automated} presented a VGG-based model for classifying benign and malignant cells from lung cytology images. \cite{noyan2020tzancknet} proposed TzanckNet based on ResNet-50 to identify cells in the cytology of erosive‑vesiculobullous diseases. Additionally, some studies compared the performance of advanced CNN architectures in cytology image classification tasks \citep{sornapudi2019comparing,hussain2020comprehensive,mohammed2021single,albuquerque2021ordinal}. From their experimental results, popular architectures (e.g., ResNet, Inception, and DenseNet) achieved promising performance in cell classification. To provide the interpretability analysis of this CNN-based classification, \citep{selvaraju2017grad} designed Grad-CAM to show region of interests (ROIs) in network decision-making using the gradient information of the last convolution layer of CNN. \cite{teramoto2019automated} utilized Grad-CAM to generate heatmaps for observing high activation areas on typical regions lung cytology images. By observing the model's high response regions of urothelial cytology images via Grad-CAM, \cite{nojima2021deep} concluded that the color of tumor nuclei contributes to the prediction of the model most. 

Apart from binary classification (i.e., benign and malignant), multi-class scenarios are more common, yet challenging in cytology image analysis, because benign and malignant cells mainly include several sub-categories \citep{vaickus2019automating}. For example, Herlev dataset contains 3 types of normal cervical cells and 4 types of abnormal cervical cells \citep{jantzen2005pap}. However, the boundaries between the image features of two sub-categories are usually ambiguous, which brings challenges for CNNs to learn distinguishable features. To solve these issues, \cite{lin2019fine} proposed a fine-grained classification model for cervical cells. This model introduced mask maps as the morphological appearance information for enhancing fine-grained distinguishable features of cells. 

In addition, several cytological studies focus on improving the model's performance in limited or imbalanced datasets. GAN-based models can be utilized to augment original dataset for improving the performance of the classification task. For example, \cite{yu2021generative} adopted GAN-based data augmentation to improve cervical cell classification models. \cite{dey2019syncgan} synthesized 180 images by conditional GAN. Together with the original data, these images are utilized to train three common models (ResNet-152, DenseNet-161, and Inception-V3), achieving significant improvement in FNAC image classification. For the imbalanced dataset problem, the number of positive samples is always far less than the negative ones in cytological scenarios \citep{yu2021generative,bakht2020thyroid}. A few studies employed sampling techniques to balance different classes \citep{li2019detection,bakht2020thyroid}. For learning-based solutions, \cite{yu2021generative} adopted GAN to balance different classes by synthesizing images for those classes with far less amount than others.

Recently, some other advanced methods have been investigated for cytology classification. \cite{li2022cervical} introduced an attention mechanism block to guide the network to focus on cell areas, thus improving the capability of extracting deep features. Then, they added a pyramid pooling layer and a long short-term memory module (LSTM) to aggregate image features in different regions. To improve the classification performance, \cite{shi2021cervical} proposed a cervical cell classification method based on graph convolutional network (GCN), which can explore the potential relationship of cervical cell images. 

In clinic practice, DL-based classification approaches have been widely applied for various types of cancers, including cervix \citep{shanthi2019deep}, breast \citep{miselis2019deep}, lung \citep{teramoto2017automated}, thyroid \citep{bakht2020thyroid}, urine \citep{vaickus2019automating}, nasal \citep{dimauro2019nasal} and skin \citep{noyan2020tzancknet}. Specifically, \cite{vaickus2019automating} proposed a VGG-based model for classifying urine cytopathology images. \cite{bakht2020thyroid} developed a VGG-based model for thyroid nodule cell classification. For cervical cytology, \cite{bao2020artificial} compared AI-assisted techniques with skilled cytologists in detecting cervical intraepithelial neoplasia or invasive cancer. For skin cytology, \cite{noyan2020tzancknet} proposed a ResNet-based model to identify cells in the cytology of erosive‑vesiculobullous diseases. These studies demonstrated the substantial clinical value of classification-assisted cytology image analysis.

\subsubsection{Slide-level classification}
Different from cell-level classification, the goal of the slide-level classification model is to predict the category of whole images instead of cell samples. 

Giga-pixel WSI classification systems are being investigated for efficient and high-accuracy predictions. Some studies built slide-level classification systems by multi-stage designs. For example, \cite{cheng2021robust} designed a robust and progressive WSI analysis method for cervical cancer screening. In the first stage, the authors developed a progressive lesion cell recognition method combining low- and high-resolution WSIs. Then, a RNN-based WSI classification model was built for WSI-level predictions in the second stage. In another slide-level study, \cite{wei2021efficient} designed a lightweight model (YOLCO) based on YOLO series \citep{redmon2016you} to make local predictions (e.g., cell-level, patch-level) in the first stage, which can enrich the multi-scale connectivity by additional supervision of spatial information. In the second stage, these local predictions were input to a transformer architecture for WSI-level results. Its experimental results showed that the framework presented a higher AUC score and $2.51\times$ faster than the state-of-the-art methods in WSI classification. For accurate and efficient screening of cervical cancer, \cite{zhu2021hybrid} developed a complete cervical LBC smear TBS diagnostic system. This system integrated XGBoost and a logical decision tree with three typical DL models, i.e., Xception for classification, YOLOv3 for object detection, and U-Net for segmentation. This diagnostic system can reduce cytologists' workload, improve the accuracy of cervical cancer screening.

Weakly supervised learning strategies are introduced to learn information from limited annotations in slide-level classification. Weakly supervised learning is appealing for this scenario. For example, \cite{dov2021weakly} presented a MIL model for thyroid cancer malignancy prediction from cytopathology images. Then, an attention module was integrated into this MIL-based model with maximum likelihood estimation (MLE) architecture. The experimental results showed the competitive performance in thyroid malignancy prediction. \cite{li2021mixed} developed mixed supervision learning for WSI classification by effectively utilizing their various labels (e.g., sufficient image-level coarse annotations and a few pixel-level fine labels).

By introducing advanced strategies or designs, quite a few cytology studies investigated to improve classification performance. For example, \cite{cao2021novel} integrated the attention module into multi-scale region-based CNN (feature pyramid network) between upsampling and downsampling pathway. Three experiments in different levels consisting of cell-level detection, patch-level, and case-level classification demonstrated the effectiveness of the introduced attention mechanism. Other studies designed different auxiliary tasks (e.g., detection, segmentation) to assist the classification task. \cite{tan2021automatic} employed an object detection model (Faster R-CNN) to assist classification for cancer screening. \cite{ke2021quantitative} introduced U-Net for improving the classification of squamous cell abnormalities.

Risk stratification is one important task of slide-level classification, which determines the risk level of patients suffering from diseases. \cite{awan2021deep} designed a DL-based digital cell profile for risk stratification of urine cytology images. In this system, RetinaNet was adopted for cell-level classification and detection in the first stage. For WSI-level risk stratification, they identified low-risk and high-risk cases using the count of atypical cells and the total count of atypical and malignant cells. \cite{lin2021dual} presented a dual-path network for cervix risk stratification, which can be divided into two steps. Firstly, an efficient CNN with a dual-path encoder was proposed for lesion retrieval, which can ensure the inference efficiency and sensitivity on both tiny and large lesions. Then, a smear-level classifier (rule-based risk stratification) was introduced to align reasonably with the intricate cytological definition of the classes. Extensive experiments on a huge dataset consisting of 19,303 WSIs from multiple medical centers validated the robustness of this risk stratification method.

\subsection{Detection}
In cytology image analysis, developing automatic detection methods to find tiny objects (e.g., malignant cells and nuclei) in the whole image is crucial to reduce experts' tedious and time-consuming workflow. DL-based object detection has achieved significant progress in medical image analysis, which can be divided into two categories: 1) One-stage method, which directly regresses the category and location of instance objects in a single architecture. 2) Two-stage method, which firstly predicts object candidates in the first stage and then classifies and localizes them in the second stage.
\begin{table*}[!t]
\scriptsize
\centering
\begin{center}
\caption{Overview of deep learning-based detection studies for computational cytology}
\label{Table_4} 
\begin{tabular}{p{1.5cm}p{2.5cm}p{0.75cm}p{0.75cm}p{3cm}p{3cm}p{3.5cm}<{\raggedright}}
\toprule

Reference & \multicolumn{1}{l}{Application} & \multicolumn{1}{l}{Staining} & \multicolumn{1}{l}{Organ} & \multicolumn{1}{l}{Method} & \multicolumn{1}{l}{Dataset}& \multicolumn{1}{l}{Result}\\ \midrule 

\multicolumn{2}{l}{\textbf{One-stage methods} }&&&&& \\ \midrule 

\cite{kilic2019automated}&Nuclei detection&Pap&Pleural effusion&YOLOv3&Private dataset: 200 images with 11,157 nuclei&Precision: 0.941, Recall=0.9898, F1=0.9648, Test time=0.060 sec/img.\\ \midrule

\cite{xiang2020novel}&Automation-assisted cervical cancer reading&Feulgen&Cervix&YOLOv3&Private dataset: 12,909 images with 58,995 ground truth boxes in 10 categories&Detection: mAP=0.602; Classification: Sens=0.975, Spec=0.687.\\ \midrule

\cite{moosavi2021histogram}&Hematological diagnosis&Giemsa&Bone marrow&YOLOv4&Private dataset: 75,000 annotated tiles of bone marrow aspirate&Region detection: Acc=0.97, AUC=0.99; Cell detection: mAP=0.75, F1-score=0.78.\\ \midrule

\cite{liang2021global}&Cell detection&Pap&Cervix&Global context-aware + Soft scale anchor matching&Private dataset: 12,909 cervical images with 58,995 ground truth boxes corresponding to 10 categories objects&mAP=0.6544.\\ \midrule

\multicolumn{2}{l}{\textbf{Two-stage methods} }&&&&& \\ \midrule 

\cite{li2019detection}&Cell detection and classification&Pap&Cervix&Faster R-CNN + Transfer learning&Private dataset: 680 LBC cervical exfoliated cell samples&Classification: Acc=0.9161 Detection: mAP=0.6698. \\ \midrule

\cite{hossain2019renal}&Nuclei detection, Estimate proliferation rate&H\&E&Kidney&R-CNN&Private dataset: 16,905 segmented cancer cell and 22,948 normal cell nuclei&P=0.9901, R=0.9870, F1=0.988.\\ \midrule

\cite{pirovano2021computer}&Localization and detection of abnormalities &Pap&Cervix&Weakly supervised CNN + Regression constraint&Herlev&Severity classification: Acc=0.952;  Normal/abnormal classification: Acc=0.952, KAPPA score=0.870; Detection: Acc=0.804.\\ \midrule 

\cite{chai2021deep}&Cell detection&Pap&Cervix&Faster R-CNN + Deep metric learning&Private dataset: 240,860 images& 100\% labeled: mAP=0.27; 75\% labeled: mAP=0.254; 50\% labeled: mAP=0.195.\\ \midrule

\cite{su2020development}&Ascites cytopathology interpretation&Pap, H\&E&Stomach&Classification: pre-trained AlexNet, VGG-16, GooleNet, ResNet18, and ResNet-50. Detection: Faster R-CNN&Ascites  2020&Classification: AUC=88.51 (ResNet50); Detection: IoU=0.8722, mAP=0.8316.\\ \midrule

\cite{xie2018efficient}&Cell detection&H\&E & Cervix&Fully residual CNN + Structured regression&HeLa cervical cancer&Precision=0.98, Recall=0.98, F1=0.98.	\\ \midrule 

\cite{marzahl2020deep}&Quantification of pulmonary hemosiderophages&Prussian turnbull&Lung&ResNet-18, FPN&Private dataset: 17 WSIs with 78,047 hemosiderophages&Concordance=0.85, mAP=0.66.\\ \midrule

\cite{zhang2019dccl}&Cervical cytology analysis&Pap&Cervix&Lesion cell detection: Faster R-CNN and RetinaNet. Cell type classification: Inception-v3, ResNet-101, and DenseNet-121&Private dataset: 1,167 WSIs with 14,432 image patches, and 27,972 labeled lesion cells&Detection: mAP=0.2116 (Faster R-CNN); Classification: Acc=0.8884, F1=0.5996 (DenseNet-121).\\ \midrule

\cite{liang2021comparison}&Cervical cancer screening (cell/clumps detection) &Pap&Cervix&Faster R-CNN + Few-shot learning + Prototype representation&CDetector&mAP=0.488.\\ \midrule

\cite{baykal2020modern}&Nuclei detection&Pap&Pleural effusion&Detector: Faster R-CNN, R-FCN and SSD &Private dataset: 200 images (11,157 nuclei)&Faster R-CNN (ResNet-101): F1=0.9812.\\

\bottomrule

\end{tabular}
\end{center}
\end{table*}

\subsubsection{One-stage methods}
One-stage algorithms detect objects by directly generating the category and coordinates of objects with the advantages of high detection efficiency, such as SSD \citep{liu2016ssd}, YOLO \citep{redmon2016you}, and RetinaNet \citep{lin2017focal}. 

Many works in cytology introduce the YOLO model as their base network due to its high efficiency. For the structure of YOLO, it divides the original image into an $S \times S$ grid cell. Then, YOLO predicts bounding boxes, confidence for each cell. Afterwards, redundant boxes are removed by the confidence threshold and non-maximum suppression. \cite{xiang2020novel} used YOLO as their detector for cervical cells. Similarly, \cite{kilic2019automated} adopted YOLO to detect nuclei in pleural effusion cytology. The authors compared the detection efficiency between one-stage and two-stage detectors \citep{ren2015faster}. Its experimental result showed that YOLO achieved a test speed of 0.060 second/image that was much faster than 1.627 second/image in Faster R-CNN (two-stage detector). Besides, \cite{moosavi2021histogram} applied YOLO on selected appropriate ROI tiles to automatically detect and classify bone marrow cellular and non-cellular objects. To improve the performance of YOLO in cervical cell detection, \cite{liang2021global} proposed a global context-aware framework by introducing an image-level classification branch and a weighted loss that can filter false positive predictions.

To improve the feature extractor for learning multi-scale features, RetinaNet was proposed by using feature pyramid network (FPN) as its feature extractor, which achieved the state-of-the-art detection performance \citep{lin2017feature}. \cite{marzahl2020deep} employed RetinaNet for generating rich and multi-scale features for functional head (e.g., box, regression, and classification). These results contributed to the quantification of pulmonary hemosiderophages in this work.

\subsubsection{Two-stage methods}
Two-stage methods use different region proposal strategies to generate bounding boxes, such as sliding windows \citep{girshick2015fast}, selective search \citep{van2011segmentation}, and region proposal network \citep{ren2015faster}. For example, Fast R-CNN designed selective search strategy to generate bounding boxes. Then, the ROI pooling layer extracts the features of each ROI. Fast R-CNN outputs softmax probabilities and per-class bounding-box regression offsets with a multi-task loss \citep{girshick2015fast}. To integrate different modules and increase the speed \citep{ren2015faster}, Faster R-CNN improves Fast R-CNN by integrating feature extraction, proposal, bounding box regression, and classification. It designs region proposal networks (RPN), which uses bounding box regression for accurate region proposal. 

To detect cell objects in the whole cytology image, some studies employed Faster R-CNN as their base architecture. For example, \cite{li2019detection} utilized Faster R-CNN to detect cervical exfoliated cells on the LBC dataset. Similarly, Faster R-CNN was also used to detect tumor cells for further classification, which formed an ascites cytopathology image interpretation system \citep{su2020development}. 

For different cytological scenarios, researchers improved detection performance by modifying architectures or integrating with other strategies \citep{hossain2019renal,marzahl2020deep}. For efficient cell and robust detection, \cite{xie2018efficient} presented a structured regression model based on a proposed fully residual CNN. This model produced a dense proximity map that exhibited higher responses at locations near cell center. Then, training this model only required annotations of the dot instead of the traditional box, which can improve efficiency of annotating. Several studies paid attention to weakly supervised learning settings in cytological detection. For example, \cite{pirovano2021computer} proposed a computer-aided diagnosis tool for cervical cancer screening. In this method, the authors designed a weakly supervised localization strategy, which performed the Integrated Gradient method \citep{sundararajan2017axiomatic} to compute attribution maps and morphological operations to obtain the localization boxes. In another work, \cite{chai2021deep} proposed a semi-supervised deep metric learning method to improve intra-class feature compactness for cervical cancer cell detection. This model learned an embedding metric space and conducted dual alignment of semantic features on both the proposal and prototype levels. From their quantitative experiments, detection of cervical cancer cell can be a challenging study, especially for some cell classes, like ASC-US.

The clinic practice not only requires high detection accuracy but also efficiency because faster detection speed is more suitable for large-scale screening scenarios \citep{lin2019fast}. Detection models usually face trade-offs between the accuracy and the speed. For example, two-stage detectors (e.g., Faster R-CNN) can achieve higher detection results while one-stage detectors (e.g., YOLO) have advantages in faster detection speed. In cytological studies, \cite{zhang2019dccl} compared the performance between two-stage (Faster R-CNN) and one-stage (RetinaNet) methods for the detection of cervical lesion cells. The results showed that the former one achieved better experimental results in average precision. In another work \citep{liang2021comparison}, the authors improved Faster R-CNN and compared it with baseline and RetinaNet in a limited data scenario. The results in cervical cell/clumps detection showed that RetinaNet (one-stage) achieved significantly faster speed (FPS). \cite{baykal2020modern} compared three detectors, i.e., Faster R-CNN (two-stage), R-FCN (two-stage), and SSD (one-stage). As a result, R-FCN achieved a higher mAP score while SSD spent less time when testing. Their experimental results validated the trade-offs of these detection models between speed and accuracy.

\subsection{Segmentation} \label{segmentation}
The segmentation task aims at morphologically delineating the object contour. For segmentation models, they assign each pixel of the image to a specific category, so it can be regarded as a pixel-wise classification task. In cytological screening, segmentation is an essential step for different applications, including 1) separating cells/clump and background from specimens, 2) morphologically distinguishing cell types, 3) accurately segmenting cellular structures, such as nuclei and cytoplasm. 

The main challenge of cytology segmentation is accurately segmenting overlapping areas between cells \citep{lu2015improved,lu2016evaluation}. To address this issue, there are mainly two solution schemes. One is dividing the cytological segmentation task into two stages. The first stage is to utilize a semantic segmentation model (e.g., U-Net) for a coarse result, followed by a series of refinement designs for overlapping areas, thus obtaining the final accurate segmentation result. The other is based on the detect-then-segment paradigm (e.g., Mask R-CNN), which detects cytology objects in whole images and output a segmentation map by mask prediction head. This type of approach can segment objects from each detected instance in an end-to-end architecture without any refinement design. Therefore, we divide the solutions of cytology segmentation into two categories: 1) segment-then-refine method, 2) detect-then-segment method.

\begin{table*}[!t]
\scriptsize
\centering
\begin{center}
\caption{Overview of deep learning-based segmentation studies for computational cytology}
\label{Table_5} 
\begin{tabular}{p{1.5cm}p{2.5cm}p{0.75cm}p{0.75cm}p{3cm}p{3cm}p{3.5cm}<{\raggedright}}
\toprule

Reference & \multicolumn{1}{l}{Application} & \multicolumn{1}{l}{Staining} & \multicolumn{1}{l}{Organ} & \multicolumn{1}{l}{Method} & \multicolumn{1}{l}{Dataset}& \multicolumn{1}{l}{Result}\\ \midrule 

\multicolumn{2}{l}{\textbf{Segment-then-refine methods} }&&&&& \\ \midrule 

\cite{falk2019u} & Cell counting, detection, and morphometry& \tabincell{l}{Fluore-\\scence} & Various & U-Net & ISBI cell tracking 2015 &IoU=0.9203 (PhC-U373), IoU=0.7756 (DIC-HeLa). \\ \midrule 

\cite{matias2021segmentation}&Segmentation, detection, and classification of cell nuclei &Pap&Oral&Classification: ResNet-34. Detection: Faster R-CNN. Segmentation: U-Net&Oral 2021& Classification: Acc=0.88, F1=0.86; Detection: IoU=0.5832; Segmentation: IoU=0.4607. \ \\ \midrule

\cite{song2014deep}&Segmentation of cytoplasm and nuclei&H\&E&Cervix&CNN+ Coarse to fine segmentation&Private dataset: 53 slides with 1400 cells&Nuclei region detection: Acc=0.9450, F1=0.9453; Segmentation: F1=0.8951±0.0215. \\ \midrule

\cite{song2015accurate}&	Segmentation of cytoplasm and Nuclei&H\&E&Cervix&Multi-scale CNN + Graph partitioning + Touching cell splitting&Private dataset: 53 images (slide)&Cytoplasm: Dice=0.95; Nuclei: Dice=0.99.	\\ \midrule

\cite{song2016accurate}&Cell segmentation&Pap H\&E&Cervix&Multi-scale CNN + Dynamic multi-template deformation&ISBI 2015. Private dataset: 21 images (each image has 30$\sim$80 cells)&	ISBI 2015: Dice=0.89; Private dataset: Dice=0.84.\\ \midrule

\cite{araujo2019deep}&Cell image segmentation and ranking&Pap&Cervix&CNN&BHS 2019&Segmentation: P=0.73, R=0.65, F1=0.69, Time=4.75s; Ranking: mAP=0.936.\\ \midrule

\cite{kowal2020cell}&Cell nuclei segmentation&H\&E&Breast&CNN + Seeded watershed&Public dataset: 80 images&Benign: Hausdorff distance=0.840, Jaccard distance=0.776;  Malignant: Hausdorff distance=0.781, Jaccard distance=0.732.\\ \midrule

\cite{bohm2019isoo}&Semantic instance segmentation of touching and overlapping objects&Pap&Cervix&U-Net&OSC-ISBI&Dice=0.895±.0.079.	\\ \midrule

\cite{zhang2020polar}	&	Cervical cell segmentation	&Pap	&Cervix	&Attention mechanism + U-Net + Random walk	& ISBI 2014 	&Nuclei: $P_{p}$=0.94 ±0.06, $R_{p}$=0.95 ±0.05, Dice=0.93 ±0.04; Cytoplasm: $TP_{p}$=0.94 ±0.06, $FP_{p}$=0.003 ±0.004, Dice=0.93 ±0.07.\\ \midrule

\cite{walter2021multistar}&Instance segmentation&Pap&Cervix&U-Net + Star-convex polygons&OSC-ISBI&Dice=0.85 ± 0.07.\\ \midrule

\cite{hussain2020shape}&Segmentation and classification of cervical nuclei&Pap&Cervix&U-Net&Herlev&Classification: Acc=0.988; Segmentation: ZSI=0.97.\\ \midrule

\cite{tareef2017optimizing}&Cytological examination (overlapping cell segmentation)&	Pap	&Cervix	&CNN+ Shape prior (dynamic shape modeling)&ISBI 2014&Nuclei: $P_{p}$=0.94 ±0.06, $R_{p}$=0.95 ±0.06, ZSI=0.94 ±0.04; Cytoplasm: ZSI=0.90 ±0.08.	\\ \midrule 

\cite{song2020shape}&	Overlapping cytoplasms segmentation&Pap H\&E&Cervix&Shape mask generator + Refining shape priors&ISBI 2015. Private dataset: 160 clumps with 962 cytoplasms&Pap: Dice=0.854 ± 0.049; H\&E:  Dice=0.846 ± 0.054.\\ \midrule

\cite{wan2019accurate}&Nuclei detection, cytoplasm segmentation&Pap&Cervix&CNN + Double-window + Image processing + Deeplab V2 + CRFs + Cell boundary refinement&ISBI 2014; ISBI 2015; Private dataset: 580 image (patch)&ISBI 2014: Dice=0.93 ± 0.04; ISBI 2015: Dice=0.92 ± 0.05; Private: Dice=0.92 ± 0.04.\\ \midrule

\multicolumn{2}{l}{\textbf{Detect-then-segment methods} }&&&&& \\ \midrule 

\cite{sompawong2019automated}&Pap smear cancer screening&Pap&Cervix&Mask R-CNN + Fine-tuning&Private dataset: 178 images in classes of normal (2,734 ), atypical (494), low-grade (148), and high-grade cells (84)&Image-level: mAP=0.578, Acc=0.917, Sens=0.917,Spec=0.917; Nucleus: Acc=0.898, Sens=0.725, Spec=0.943.\\ \midrule

\cite{zhou2019irnet}& Cell segmentation&Pap&Cervix&PRN + Cell association matrix&Private dataset: 413 images (annotated 4,439 cytoplasm and 4,789 nuclei)&Cytoplasm: AJI=0.7185, F1=0.7497; Nuclei: AJI=0.5496, F1=0.7554.		\\ \midrule

\cite{zhou2020deep}&Cell instance segmentation&Pap&Cervix&RPN + Knowledge distillation&Private dataset: 413 labeled (4,439 cytoplasm and 4,789 nuclei) and 4,371 unlabeled images&100\% labeled: AJI=0.6643, mAP=40.52; 80\%. labeled: AJI=0.6692, mAP=0.4013; 40\% labeled: AJI=0.6449, mAP=0.3726.	\\

\bottomrule
\end{tabular}
\end{center}
\end{table*}

\subsubsection{Segment-then-refine method}
Cytological structures can be segmented by segmentation models, like U-Net. However, overlapping areas belonging to several cells bring defiance of accurately segmenting each cell structure. To overcome this issue, different refinement strategies are proposed to add after the coarse segmentation model for fine-level segmentation.

The segmentation network was originally implemented by establishing a pixel-wise classification network through CNN. \cite{song2015accurate} designed a multi-scale convolutional network for coarse segmentation. \cite{kowal2020cell} proposed a more complex CNN structure consisting of four convolutional layers, two max-pooling layers, and one fully connected layer. This architecture was utilized in the first stage for nuclei segmentation in cytological images.

Afterwards, U-Net almost replaces the previous pixel-level classification network after its occurrence, especially in biomedical image segmentation \citep{ronneberger2015u}. U-Net is a downsampling-upsampling structure with skip connections for combining low-level and high-level features. Recently, U-Net has made great achievements in medical image segmentation. For example, \cite{falk2019u} designed U-Net for cell counting, detection, and segmentation. This work illustrated its potential for cellular structure analysis. In cytology image segmentation, U-Net has been introduced as the backbone for segmenting cellular objects in various cytology, such as oral \citep{matias2021segmentation}, cervix \citep{araujo2019deep}, and breast \citep{kowal2020cell}. Several works focus on improving the performance of U-Net to enhance their capacities of cell segmentation. For instance, \cite{bohm2019isoo} proposed to mix 2D and 3D U-Net for semantic instance segmentation of touching objects. \cite{zhang2020polar} introduced attention mechanism to improve U-Net for focusing on ROIs. Besides, \cite{hussain2020shape} improved U-Net by adding residual blocks, densely connected blocks, and a fully convolutional layer as a bottleneck between encoder-decoder blocks for nuclei segmentation in cervical images.

In segment-then-refine methods, the second stage is to address the issue of overlapping and refine segmentation results. Most of them take the shape prior of cell into considerations. For instance, \cite{song2016accurate} proposed a dynamic multi-template deformation model together with high-level morphological constrain for further boundary refinement. \cite{kowal2020cell} designed a series of refinement strategies in the second stage: conditional erosion for determining nuclei seeds, the seeded watershed for separation overlapping nuclei, and aggregating segmentation results for overlapping and non-overlapping nuclei. In addition, \cite{zhang2020polar} proposed a graph-based random walk method for extracting both nucleus and cytoplasm of overlapping cervical cells. This method utilized polar coordinate sampling for removing fake nuclei. Its experimental results in ISBI 2014 dataset showed the performance improvement on extracting an individual cell from heavy overlapping cell clumps. \cite{walter2021multistar} proposed to predict object probability, star distance, and overlap probability based on U-Net. Then, non-maximum suppression was used to generate overlapping cell segmentation results. These refinement strategies can achieve more accurate segmentation results, especially for overlapping regions. However, complex clinical data will present more challenges to the reproducibility and generalizability.

In addition, some other models have been proposed for cytological segmentation. \cite{tareef2017optimizing} designed a two-stage segmentation model, which consists of initial segmentation based on Voronoi diagram, and final segmentation with learning shape prior model. In order to segment overlapping cervical cytoplasms, \cite{song2020shape} proposed a shape mask generator to refine shape priors. \cite{wan2019accurate} presented an architecture for cell detection and cytoplasm segmentation. In this method, conditional random field algorithm (CRFs) and cell boundary refinement were utilized to achieve accurate segmentation of overlapping cells in cervical cytology. 
\begin{table*}[!h]
\scriptsize
\centering
\begin{center}
\caption{Overview of deep learning-based studies of other tasks for computational cytology}
\label{Table_6} 
\begin{tabular}{p{1.5cm}p{2.5cm}p{0.75cm}p{0.75cm}p{3cm}p{3cm}p{3.5cm}<{\raggedright}}
\toprule

Reference & \multicolumn{1}{l}{Application} & \multicolumn{1}{l}{Staining} & \multicolumn{1}{l}{Organ} & \multicolumn{1}{l}{Method} & \multicolumn{1}{l}{Dataset}& \multicolumn{1}{l}{Result}\\ \midrule 

\cite{ma2020pathsrgan}&Super resolution&Pap&Cervix&Image registration + GAN &Private dataset: 142 WSIs (118 for training and 24 for testing) with 174,500 patches&PSNR=26.92, SSIM=0.88, MOS=3.80.	\\ \midrule 

\cite{ma2021stsrnet}&Super resolution&Pap&Cervix&Backbone + Self-texture + Flexible reconstruction &Public dataset: 5 slides (25,000 patches)&PSNR=35.47, SSIM=0.958, MSE=22.07.	\\ \midrule 

\cite{teramoto2021mutual}&Mutual stain conversion&Giemsa and Pap&Lung&CycleGAN&Private dataset: 191 Giemsa-stained images and 209 Papanicolaou-stained images&T test: P-value $<$0.001.\\ 

\bottomrule
\end{tabular}
\end{center}
\end{table*}

\subsubsection{Detect-then-segment method}
For the segmentation of instance objects in cytology images, these methods follow the detect-then-segment paradigm, which divides this task into two steps: detecting all objects in whole images, then segmenting instances from each detected object. As one popular architecture of this paradigm, Mask R-CNN improves Faster R-CNN by adding full connected layers as the segmentation head. Thus, it can output the prediction of classification, detection, and segmentation via a single architecture \citep{he2017mask,ren2015faster}. 

Instance segmentation is regarded as one of the most challenging tasks in cytology image analysis, because it not only predicts the instance morphology but also distinguishes different instances (e.g., cytoplasm, nucleus). Building detect-then-segment models can solve this problem, since they can predict instance detection and segmentation results through a single architecture. A few studies investigated this category of cytological segmentation methods. Existing researches almost employ Mask R-CNN as their architecture, because there is no need for further design of overlapping areas in this category of methods. For example, \cite{sompawong2019automated} adopted Mask R-CNN for specifying the bounding box, nucleus mask, and class of each cervical cell. In another study \citep{zhou2019irnet}, authors utilized the multi-head attention mechanism to explore instance-level association by propagating features based on attention scores. Consequently, this proposed model improved the instance representation and achieved better instance segmentation performance than original Mask R-CNN. To further reduce reliance on large amounts of labeled data, \cite{zhou2020deep} proposed a semi-supervised learning approach to leverage both labeled and unlabeled data for instance segmentation by knowledge distillation.

\subsection{Other tasks}
In addition to the typical deep learning tasks, i.e., classification, detection, and segmentation, a few other tasks of cytology image analysis have also been investigated, such as super-resolution (SR), and stain conversion (Table \ref{Table_6}). In the cytopathology screening, low-resolution and out-of-focus images will harm the decision-making process of cytologists, thus high-resolution digital cytopathology slides are the prerequisite for the interpretation of lesion cells. To control the image quality, super-resolution models are designed to generate high-resolution images. \cite{ma2020pathsrgan} introduced a GAN-based progressive multi-supervised super-resolution model (PathSRGAN) to learn the mapping of real low-resolution and high-resolution images. After that, they designed a self-texture transfer super-resolution and refocusing network (STSRNet) to reconstruct HR multi-focal plane (MFP) images from a single 2D low-resolution (LR) wide filed image \citep{an2021stsrnet}. As mentioned in section \ref{Preprocessing}, different staining methods are used to observe different cell structures and components. DL-based stain conversion can be used for staining normalization and eliminate data heterogeneity issues. \cite{teramoto2021mutual} proposed a CycleGAN-based style transfer model for stain conversion between Giemsa-stained and Pap-stained images. This study performed visual evaluations of the authenticity of cell nuclei, cytoplasm, and cell layouts of synthetic lung cytology images.

\section{Challenges and Promises}\label{Promises and Challenges}
Despite great advancements and improvements in computational cytology over the last few years, there are still quite a few challenges and opening problems that are waiting to be resolved. Meanwhile, the development of deep learning technologies and pathology is continuously bringing vigor and vitality into this emerging field. In this section, we further discuss prospects and potential research directions of computational cytology.

\subsection{Label-efficient learning with limited annotations} 
Data is regarded as the prerequisite of learning-based methods, because it is hard to develop effective models without good quality datasets \citep{falk2019u,jonsson2019brain}. In medical image analysis, large-scale labeling can be a heavy burden for cytologists, because they need to first manually delineate ROIs in WSIs, then annotate each object (e.g., nucleus, cell, and cluster) in these ROIs by the box or mask. Compared with the extensive dataset in histopathology (such as TCGA \citep{tomczak2015cancer}), public datasets of cytology are more limited not only in their numbers, but also in cancer types and annotation types (see in Table \ref{Table_1}). Therefore, how to efficiently utilize datasets with limited annotations can be challenging for developing cytological analysis models \citep{dov2021weakly,chai2021deep}.

In recent years, the concept of label-efficient learning is proposed to makes full use of the limited annotations for leveraging information, including semi-supervised learning, multiple instance learning, mixed supervised learning, etc. Specifically, semi-supervised learning aims to solve this problem by learning knowledge from both labeled data and unlabeled data. Recently, \cite{zhou2020deep} designed a semi-supervised learning method by a mask-guided teacher-student framework for overlapping cell instance segmentation. Another learning scheme, MIL utilizes image-level annotations for instance-level tasks, which has been investigated in the field of medical image analysis, especially for histopathology. Recently, \cite{dov2021weakly} proposed a MIL-based algorithm in thyroid cytology, which can simultaneously predict multiple bag and instance-level labels for thyroid malignancy prediction from WSIs. However, the potential of MIL in pap smear image and other cytology images remains to be explored. As illustrated in Table \ref{Table_1}, there are usually different annotation types in different datasets, including box, mask, and the image-level label. Recently, mixed supervised learning gains popularity in analyzing images with different types of annotations. It has been demonstrated as an effective learning scheme in medical domain. For example, \cite{luo2021oxnet} present a deep omni-supervised thoracic disease detection network from chest X-rays with massive image-level annotations and scarce lesion-level annotations. This learning paradigm can substantially reduce the demand for fine annotation, thus reducing workload of doctors significantly.

In addition to the effective utilization of annotations, it is also promising to build efficient labeling approaches for reducing the burden on annotators. For example, introducing human knowledge into the loop of annotation can effectively and actively obtain accurate and credible annotations. \cite{koohbanani2020nuclick} proposed an annotating method named NuClick with a squiggle as a guiding signal, enabling it to segment the glandular boundaries. However, this method still requires human full attention to annotate samples, which is a huge cost for society and tedious for human experts. For cytology images, they always contain numerous cells, especially in giga-pixel WSIs, how to establish an effective labeling process remains largely unexplored.

\subsection{Fine-grained classification and morphological feature characterization}
Although various deep learning applications have been developed in cytology image analysis (e.g., classification, detection, and segmentation), some of these tasks are worthy of further investigation, such as fine-grained classification for cancer diagnosis and instance segmentation in overlapping cell scenarios.

As a fundamental task, cytological classification aims to distinguish between benign and malignant cells. However, there are many types of cells with varying degrees of cancerization. For example, the types of cervical cells include squamous cells and glandular cells, and they can be further subdivided into subcategories \citep{nayar2015bethesda}. Besides, each type of cell has large intra-variance and small inter-variance. These factors bring significant challenges to deep feature extractors for learning distinguishable features. Further fine-grained classification of malignant cells can solve these problems and assist cytologists in accurate cancer diagnosis. \cite{zhang2019dccl} demonstrated the difficulty of fine-grained tasks compared to coarse-grained tasks. In their experiments, the classifier of cervical cytology showed significantly worse performance of fine-grained than coarse-grained tasks. Some studies introduced additional information and prior to learn fine-grained feature representations. For example, \cite{lin2019fine} built a fine-grained classification model for distinguishing cervical cells. The authors introduced cytoplasm and nucleus masks as morphological information to extract fine-grained features. Therefore, fine-grained feature extractors for learning subtle feature diversity are essential for cytology image analysis. 

Another fundamental application, instance-level segmentation of cytoplasm and nucleus can be used to calculate the nuclear-cytoplasmic ratio, which is an essential indicator for distinguishing malignant cells. However, cell overlapping brings challenges for accurately instance segmenting cellular structures \citep{song2014deep,song2015accurate}. A large amount of studies focus on this challenge, and most of them divide this task into two stages: semantic segmentation models for coarse segmentation, followed by refinement processing technologies for overlapping areas \citep{song2016accurate,zhang2020polar}. These multiple-stage architectures introduce lots of human designs and interventions, leading to increasing training difficulties and poor generalizations. Recently, few studies investigate the feasibility of building end-to-end models by the detect-then-segment paradigm \citep{zhou2019irnet}, which detect objects and then segment each detected object in a single learning framework. Although cytological segmentation has made some progress, for complex morphological feature representation, building end-to-end instance segmentation models still remains value to be studied and explored.

\subsection{Effective feature representation learning}	
In deep learning, the performance of downstream tasks will be greatly influenced by the feature representation capability of feature extractors. The main goal of DL models in cytology is to learn effective features of cytology images for cell classification, cellular objects detection and segmentation. Thus, building models that effectively represent features is crucial for cytology image analysis.

Recently, attention mechanism has made promising achievements in effective representation of image features, thus improving performances of down-stream tasks \citep{vaswani2017attention}. The deep neural networks with attention modules can reduce the dependence on external information, and be better at capturing internal correlations of data or features. In cytology images, there are too many useless objects, like background, mucus, blood, and inflammatory cells \citep{hussain2020liquid}. Attention mechanism-based strategies can make deep models focus more on lesion-related regions or object-related (e.g., nuclei, cytoplasm) regions. Besides, the attention mechanism not only improves the deep learning model but also provides convenience for model visualization and understanding. A few studies have validated the superiority of the attention mechanism in cytology image analysis. \cite{zhang2020polar} utilized the attention mechanism to enhance U-Net for segmentation of overlapping cervical cells. Besides, \cite{zhou2019irnet} introduced a multi-head attention mechanism module into the instance segmentation model for improving instance representation. These studies integrated attention modules into the framework and achieved improved performances. After that, the popular structure with attention mechanism, transformer has been demonstrated its superiority of learning global dependencies in various applications \citep{dosovitskiy2020image}. Transformer-based structures are waiting to be investigated in cytology image analysis, such as building dependencies between different cells in patch images, or different regions in WSIs.

\subsection{Generalizability and robustness} 
The generalizability of DL-based algorithms in computational cytology determines whether the model can be successfully applied to actual clinical scenarios with various influences. At present, many DL approaches for medical image analysis can obtain acceptable performance in their own datasets, but their clinical performance is far from reaching practical standards \citep{litjens2017survey}.

Due to various specimen collection methods, staining techniques and imaging protocols, the data heterogeneity is the key reason for the poor clinical performance. It can lead to the weak robustness and generalization capability of DL models, thus performing poorly when applied to unseen data scenarios. Clinically, the performance of these DL models could degrade significantly, leading to low clinical reproducibility \citep{kowal2020cell}.

Extensive researchers are investigating to mitigate this issue. For image processing strategies, normalization methods are utilized to pre-process input data in many image analysis tasks. These methods provide limited benefit in DL-based medical image analysis, because two datasets can be influenced and normalized against each other \citep{mahmood2019deep}. For learning strategy, domain adaptation can alleviate this problem by transferring knowledge for decreasing the need for annotations of target tasks \citep{oza2021unsupervised}. Domain adaptation has been used for other medical scenarios with cross-domain data (e.g., CT and MRI \citep{chen2020unsupervised}), the potential of domain adaptation methods in cytology image analysis remains to be explored. For instance, building cross-domain learning framework to analyze multi-domain images, such as images from multi-center, differently stained cytology images, and even different pathology images (e.g., cytology and histopathology). 

Designing new models for robust feature extraction of cytology images is also worthy of being explored, especially for feature extractor, which aims to map cytology images from multi-center to the same feature space for learning domain-invariant representations, thereby addressing the data heterogeneity issue \citep{li2021mixed}. In addition, multimodal data has been demonstrated to compensate for the missing information in single modality data \citep{zhang2021deep}. For medical scenarios, \cite{shao2019integrative} designs a multimodal fusion framework to combine histopathological images and genomic sequences for early-stage cancer prognosis. Similarly, multimodality or full modality learning is also worth exploring and studying in cytology applications.

\subsection{Transparency and interpretability}	
Unlike other deep learning application scenarios, the medical domain not only focuses on the model performance in clinical practice but also its interpretability, which is of paramount importance in clinical decision-making. However, the black-box nature of DL algorithms lacking clinical interpretability and transparency has restricted its clinical adoption \citep{lu2021data}. Therefore, explainable AI for medicine is introduced to establish the confidence between AI technologies and doctors/patients. There are some explorable issues in computational cytology, such as slide-level cytology screening based on rules, and visualization of the decision-making process of deep models.

Visualization is regarded as one of techniques for interpretation. Current techniques mainly utilize attention mechanism to generate heatmaps to visualize the deep features and models, such as class activation mapping (CAM) \citep{zhou2016learning}. In cytology image analysis, a few studies have employed heat maps or feature maps for exploring the decision-making basis of deep learning models in the classification task. \cite{nojima2021deep} used Grad-CAM to observe the model's high response regions of urothelial cytology images, which can improve medical decision-making from the gradient of the differentiable model. For model-agnostic visualization, another technique, local interpretable model-agnostic explanation (LIME) \citep{ribeiro2016should} can be effective to provide the interpretability and transparency. LIME presents a locally faithful explanation by fitting a set of perturbed samples near the target sample using a potentially interpretable model. In addition to these mentioned methods, more visualization techniques are under exploration for interpretability.

To increase the credibility of deep models in computational cytology, researchers constructed slide-level screening system by assembling multiple tasks rather than predicting the final diagnosis results of cytology slides, which can assist pathologists to obtain multi-stage analysis results. \cite{zhu2021hybrid} integrated DL-based classification, detection, and segmentation models to build a cervical LBC smear TBS diagnostic system. Another study, \cite{lin2021dual} built a dual-path network for outputting the detected lesions, followed by a rule-based risk stratification system. \cite{wei2021efficient} divided the cervical WSI analysis into two stages, i.e., cervical lesion detection at the patch level in preselected ROIs, and normal/abnormal classification at the WSI level. The output of each stage of these multi-stage methods can provide intermediate explanations for model prediction and cytology screening, thus improving the transparency of the diagnostic process.

\subsection{Digital medicine and human-AI collaboration} 
Digital medicine integrates medicine and information technology for clinical diagnosis and treatment, it aims to transform digital models to clinical scenarios \citep{beam2020challenges,samuel2020machine}. Although a large number of DL models report that they can achieve state-of-the-art performance, they are mostly validated on domain-specific datasets and cannot achieve the same good performance in clinical practice. Unavailability of data is one of the crucial factors leading to this problem, because data privacy cannot be overemphasized in medicine domain. Currently, privacy-preserving learning approaches are being explored and studied for improving the availability of multi-center data \citep{chen2021machine}, e.g., federated learning \citep{rieke2020future}. Another crucial issue facing clinical transformation is that clinical data is usually more diverse and complex than collected training data, caused by variable clinical factors regarding imaging microscopes, staining techniques, patch extraction, and selection, etc. To address this issue, designing more robust architectures can make the model less dependent on data quality in digital medicine. In addition, cytologists analyze specimens of different cancer types using different diagnostic criteria \citep{nayar2015bethesda}, which makes it difficult for DL algorithms that focus on domain-specific cytology to adapt to different cancer scenarios. 

In recent years, numerous studies show that the human-machine collaborative medical diagnosis system can achieve better diagnosis performance than conventional diagnosis systems by the intervention of human experts and assistance of machines \citep{zhu2021hybrid, patel2019human}. The DL algorithm in the intelligent medical system provides doctors with multi-level, and high-confidence prediction results. Besides, the digital diagnostic system provides doctors with real-time diagnostic information through human-computer interaction technology. The mart microscope system (ARM), designed by Google Health, has made an early breakthrough in this field \citep{chen2019augmented}. ARM integrates AI algorithms with optical microscope to analyze pathological slides and provide pathologists analysis results (e.g., lesion contour, probability heatmap) in the field of view by augmented reality technologies. Then, pathologists make the final diagnosis based on these quantitative and qualitative results. This computer-assisted diagnostic system has provided prospects for the automation of histopathology, cytology, parasitology, etc. However, there is still a long way to go in terms of accurate and efficient diagnosis, system integration, hardware resources, and AI algorithms.

\section{Conclusion}\label{Conclusion}
In recent years, the development of deep learning has enabled great success in computational cytology, showing signiﬁcant promise for efficient cancer screening. In this paper, we have comprehensively reviewed the current progress of deep learning-based methods in computational cytology, including supervised learning, weakly supervised learning, unsupervised learning, and transfer learning. More specifically, we survey image analysis-based approaches and state-of-the-art DL algorithms with the applications of classification, detection, and segmentation in cytology. Various applications of advanced DL-based works of various cytology were investigated in this paper, including the cervix, breast, lung, thyroid, oral, kidney, stomach, etc. We also summarize the evaluation metrics and public datasets for developing new models. Finally, we outline current challenges and potential directions for future research of computational cytology.

\section*{Declaration of Competing Interest}
The authors declare that they have no known competing financial interests or personal relationships that could have appeared to influence the work reported in this paper.

\section*{CRediT authorship contribution statement}
\textbf{Hao Jiang:} Conceptualization, Methodology, Writing - original draft, Visualization.
\textbf{Yanning Zhou:} Conceptualization, Formal analysis, Writing - review \& editing.
\textbf{Yi Lin:} Conceptualization, Writing - review \& editing, Investigation.
\textbf{Ronald CK Chan:} Formal analysis, Writing - review \& editing.
\textbf{Jiang Liu:} Conceptualization, Investigation.
\textbf{Hao Chen:} Conceptualization, Funding acquisition, Project administration, Resources, Supervision, Writing - review \& editing.

\section*{Acknowledgments}
This work was supported by Beijing Institute of Collaborative Innovation Program (No. BICI22EG01).

\bibliographystyle{model2-names.bst}\biboptions{authoryear}
\bibliography{ref}
\end{document}